\documentclass{aa}  
\usepackage{natbib}
\usepackage{graphicx}
\usepackage{verbatim} 
\usepackage{txfonts}

\begin{document}
\title{An integrated optics beam combiner for the second~generation VLTI~instruments} 

\author{M.~Benisty\inst{1},   J-P.~Berger\inst{1},   L.~Jocou\inst{1},
  P.~Labeye\inst{2}, F.~Malbet\inst{1}, K.~Perraut\inst{1} and P.~Kern\inst{1}} 
                                       
   \offprints{Myriam.Benisty@obs.ujf-grenoble.fr}

   \institute{Laboratoire d'AstrOphysique de  Grenoble (LAOG), 414 rue
              de la piscine, 38400 St Martin d'Heres, France \\
            \and
         CEA-LETI, Minatec, 17 rue des martyrs, 38054 Grenoble, France}
         
   \date{Received 06/10/2008; Accepted 28/01/2009}

  \abstract
  {Recentely, an increasing number of scientific publications making use of
images  obtained with near-infrared  long-baseline interferometry  have
been produced. The technique has reached, at last,
a technical maturity level that opens new avenues for numerous astrophysical topics requiring
milli-arc-second model-independent  imaging. The Very  Large Telescope
Interferometer (VLTI) will soon be equipped with instruments
able to combine between four and six telescopes.}  
{In the framework of the VLTI second generation instruments Gravity and VSI, we propose a new beam
  combining concept using integrated optics (IO) technologies with a novel ABCD-like fringe encoding
  scheme. Our goal is to demonstrate that IO-based combinations bring considerable advantages in
  terms of instrumental design and performance. We therefore aim at giving a full characterization
  of an IO beam combiner in order to establish  its performance and check
  its compliance with the specifications of an imaging instrument. }
{For this purpose, prototype IO beam combiners have been manufactured and laboratory measurements were
  made in the H band with a dedicated testbed, simulating a four-telescope interferometer. We
  studied the beam combiners through the analysis of throughput, instrumental visibilities, phases and closure phases
  in  wide band as  well as  with spectral  dispersion.  Study  of the
  polarization properties was also carried out. 
  } 
  {We obtain competitive throughput (65\%), high and stable instrumental contrasts (from 80\% in
    wide band up to 100\%$\pm$1\% with spectral dispersion), stable but non-zero closure phases
    (\textit{e.g.}  115$^\circ \pm$2$^\circ$)  which  we attribute  to
    internal optical path differences (OPD) that can be calibrated. We validate a new static
    and an achromatic phase shifting IO function 
    close to the nominal 90$^\circ$ value (\textit{e.g.} 80$^\circ \pm$1$^\circ$). All
    these observables show limited chromaticity over the H band range.}
  {Our results demonstrate that such ABCD-like beam combiners are
    particularly  well  suited   for  interferometric  combination  of
    multiple beams  to achieve aperture synthesis  imaging. This opens
    the  way  to  extending   this  technique  to  all  near  infrared
    wavelengths and in particular, the K band.}  

\keywords{optical interferometry -- integrated optics}

  \authorrunning{Benisty et al.}
  \titlerunning{A 4-beam IO beam combiner with ABCD encoding}
  
\maketitle
%

\begin{figure*}[t!]
  \begin{center}
    \begin{tabular}{c}
      \includegraphics[width=0.6\textwidth]{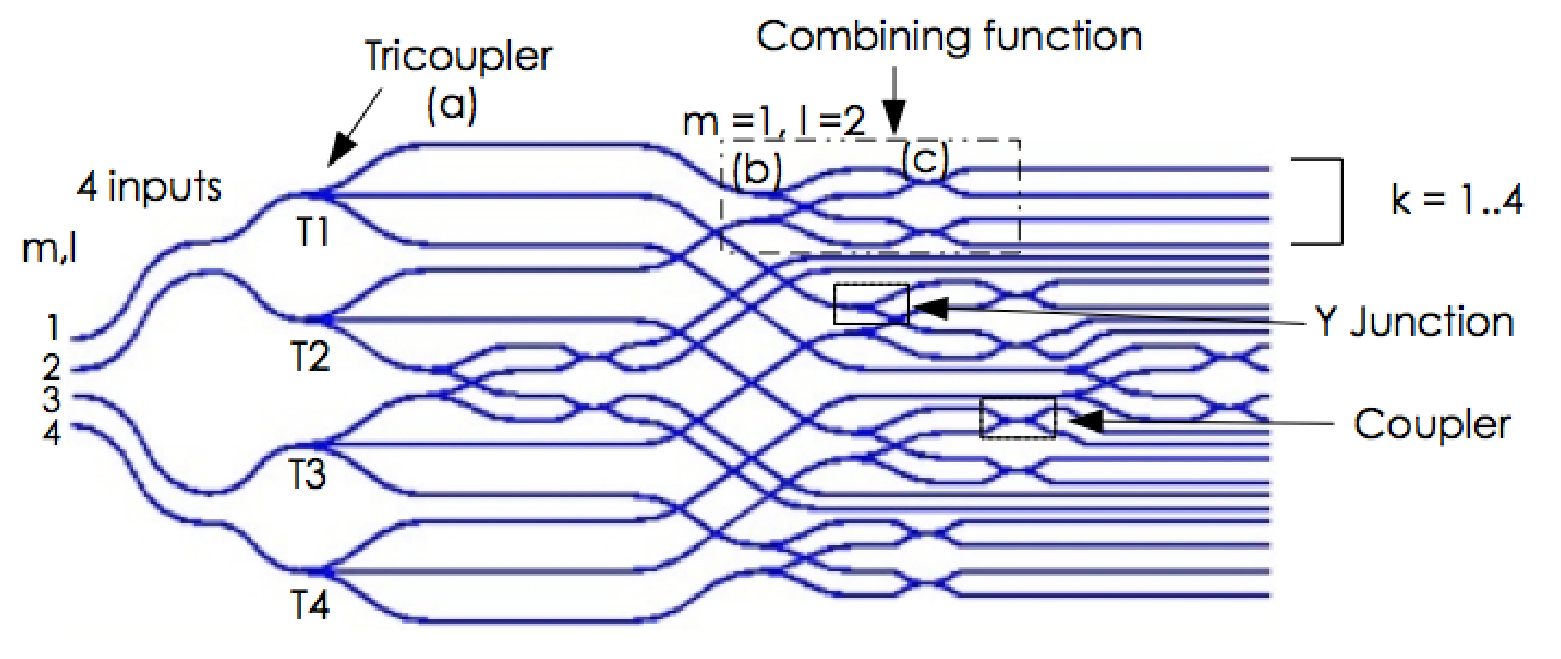} \\
      \includegraphics[width=0.55\textwidth,angle=180]{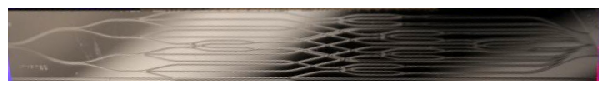} 
 \end{tabular}
    \caption{\label{fig:composant} Upper panel:  theoretical design of the
      integrated optics 4-way beam combiner allowing pairwise combination and using
      phase-shifting  devices  to  produce  4 outputs  in  quadrature.
      We  refer  to each  output  using  the  index
      \textit{m,l,k}.   $ml^{k}$  is the  k$^{th}$  output  out of  4,
      resulting from  the combination  of the beam  $m$ and  $l$.  The
      lower panel is a picture of a prototype that is 80mm long and 8mm 
      wide.}  
\end{center}
\end{figure*}

\section{Introduction}

Optical long baseline interferometry offers a unique way to directly probe astrophysical
environments  with milli-arcsecond  resolution.  The study  of
  stellar surfaces,  evolved stars,  young stars, our  galactic center
  and the heart of active galactic nuclei require access to direct imaging. Until now, a large fraction of observations in the near
infrared (NIR) were obtained with 2 to 3-telescope arrays, with little spatial frequency coverage
(so-called \textit{uv} coverage), restricting the astrophysical interpretation to a parametric one
in most of the cases. However, discriminating between different successful scenarios of complex or
rapidly-changing objects raises the need for images as model-independent as possible. This
translates into the requirement to use as many telescopes as possible in order to fill the
\textit{uv} plane and allow an unambiguous image reconstruction.  Until very recently, most of the
images produced with optical long baseline interferometers had moderate complexity and therefore did
not bring additional information with respect to parametric modelling. In our opinion, the
complexity barrier where the reconstructed image adds meaningful scientific value to the
astrophysical interpretation was recently passed by \citet{Monnier:2007} and \citet{Zhao:2008} using
the MIRC instrument, an image plane 4-beam combiner using single mode
  fibers, at the CHARA interferometer \citep{mirc,chara05}. 

The  Very Large  Telescope Interferometer  (VLTI, \citep{vlti1,vlti2})
will be equipped in 2012-2015 with two second-generation instruments: Gravity 
\citep{gravity} and VSI \citep{vsi2,malbet08} which will be capable of exploiting the imaging capability of the array by
combining four beams for the first and six for the second. The stringent requirements for these two
instruments have triggered the interest in using integrated optics (IO) as a core technology for the beam
combining function.  The ability  to integrate a singlemode circuit on
a substrate, able to interfere all the beams offers numerous advantages both in terms of performance and ease of operation. 
Single-mode beam combiners provide natural modal filtering, which associated
with proper photometric calibration has been shown to lead to accurate visibility measurements. The
compactness of the chip allows the instrument footprint to be minimized and the thermal control to
be  optimized   (further  enhancing  the   calibration  accuracy).  No
alignment is required,  other than the injection in  the input guides,
even though the combination scheme is complex. Finally, this technology offers the
flexibility to easily  switch beam combiners to adapt  to a particular
situation (\textit{e.g.} target, number of telescopes).  

Since the initial proposition by \cite{kern96}, LAOG and its industrial partner LETI/CEA
have been developing the use of IO technology to interferometrically combine light beams in optical
waveguides lying on a solid substrate of a few centimeter \citep{kern96,malbet99,berger00}. This
instrumental research program has consisted in designing, fabricating and characterizing all the IO
building  blocks  required to  build  an astronomical  interferometric
beam  combiner.  Several  beam  combining schemes  have  been
implemented and  tested. Some of  them have led to  successful on-sky
demonstrations such  as the VINCI/VLTI (2  telescopes) and IONIC3/IOTA
(3 telescopes) instruments \citep{berger03,lebouquin04,kraus05,monnier06}. 

In the context of VLTI second-generation instrument studies, \cite{lebouquin05} have studied the
global efficiency of a great  variety of IO beam combiners. This study
has concluded that one of the most efficient ways to combine four beams (\textit{e.g.} 4 UT or 4 AT) was to use a so-called ``pairwise static
ABCD''   scheme    (inspired   by   the    visibility   estimator   of
\citet{Shao:1977}).   This   IO   circuit   allows  one   to   extract
simultaneously four phase states  of the coherent signal independently
for each of the six 
baselines. We fabricated them \citep{labeye08}, and in this paper, we present these new 4-beam
combiners     together      with     their     complete     laboratory
characterization.   They   are   probably   the   most   sophisticated
astronomical beam  combiners built  to  date.  The paper  is
organized as follow: in Sect.~2, the technology 
and the specific design of the beam combiners are described.  In Sect.~3,
we  present  the  laboratory  set  up  as  well  as  the  experimental
procedure;  the characterization results  are given  in Sect.~4 and
discussed in Sect.~5.


\begin{figure}[t]
  \begin{center}
    \begin{tabular}{cc}
      \includegraphics[width=0.5\textwidth]{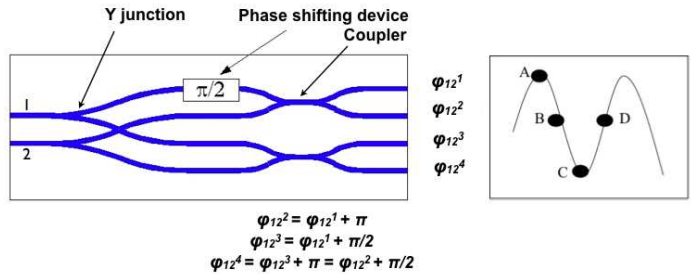}
   \end{tabular}
 \end{center}
 \caption
 { \label{fig:abcd} Details of the beam combining function: for each
   interferometric  pair  (\textit{e.g.}   [12]), one  arm  is
   shifted  by   90$^\circ$    leading   to    four    outputs   in
   quadrature  (with phases  written as  $\varphi_{12}^{1}$ to
   $\varphi_{12}^{4}$). Combinations of  beams occur in couplers that
   present  two  outputs  in   phase  opposition  to  maintain  energy
   conservation. By  recording the  four phase states  (ABCD-like, see
   the right figure), one can retrieve the interferometric observables
   (amplitude and phase of the fringes).}
\end{figure}

\section{The beam combiner: technology and design}
\label{sec:tech}

Prior to fabrication, the IO circuit was designed and numerical
computation  simulating the propagation  of an  electromagnetic signal
was carried out to determine  the expected properties in terms of flux
routing. Each IO function was checked and its throughput and flux 
distribution   were  optimized  numerically.   This  step   done,  the
simulation parameters  were turned into technological  parameters and a
photolithographic mask was fabricated.  

LETI uses a silica-on-silicon technology to fabricate IO circuits. This technological process
requires several photolithographic steps to etch different layers.
The beam combiners are made by
depositing alternatively 3 doped silica layers on a silicon substrate. The second layer is etched to
define channel waveguides and the other two layers constitute the optical cladding. For
the first time, the etching technology allows us to completely isolate each waveguide from the others
\citep{labeye06}. The produced beam combiners have been designed to operate
in the atmospheric H~band and more recently in the K~band.

\begin{figure}[t]
  \begin{center}
    \includegraphics[width=0.35\textwidth]{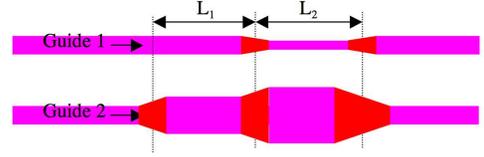}
    \caption{\label{fig:dephaseur}  Principle  of  the  phase-shifting
      function:  a  variation  of  the  optical path  is  induced  by  a
      differential change of the waveguide effective index due to a
      change in their width.  The concatenation of carefully optimized portions of waveguides with
      controlled effective index allows to flatten the wavelength response.} 
  \end{center}
\end{figure}

The so-called  ``pairwise static ABCD'' beam  combiner can be
described as follows.  Each beam combiner is designed
  to have 4 inputs and 24 outputs, allowing 6 interferometric pairwise
  combinations, each one producing 4 phase-shifted outputs with a
  phase difference of 90$^\circ$.  For each injected beam, the light propagates through waveguides and is split in three
in a  tricoupler (item (a)  in Fig.~\ref{fig:composant}) to  enter the
combining function (constituted of Y-junctions and couplers).  The light
is then divided in two in a Y junction that acts like a classical beamsplitter (item b), each
beam later  being combined in  a coupler (item  c) with a  beam coming
from another  telescope. A coupler allows a  controlled power transfer
between one waveguide and another. As a consequence of 
energy   conservation,  each   coupler  has   two  outputs   in  phase
opposition.   In only  one  of  these four  arms,  there is  a
  phase-shifting  device   designed  to  change  the   phase  of  the
propagating beam  by 90$^\circ$ (Fig.~\ref{fig:abcd}).   This leads to
four  output beams,  two of  them being  in phase  opposition  with an
additional  phase-shift of 90$^\circ$  with respect  to the  other two
(\textit{e.g.}  $\varphi_{12}$ and $\varphi_{12}$+$\pi$; $\varphi_{12}$+$\pi/2$
and $\varphi_{12}$+$\pi$+$\pi/2$, following Fig.~\ref{fig:abcd}).  
 The  phase-shifting  function  is  based  on  the  variation  of  the
 effective index  (\textit{i.e.}  index  seen by the  fundamental mode
 propagating   into  the  waveguide)   with  the   waveguide  diameter
 \citep{labeye08}.  To create a phase shift, enlarging one of 
 the two waveguides creates a difference in the effective index and leads to an \textit{optical}
 path difference between two parallel waveguides of the same
 \textit{physical} length. In order to achieve an achromatic phase shift,
 the wavelength dependence is compensated by concatenating a few
 waveguide  segments  of   different  diameters  separated  by  tapers
 (\textit{i.e.}   adiabatic  functions)  to  avoid  any  loss  due  to
 discontinuities   (Fig.~\ref{fig:dephaseur}).   Since the  photometry   is
 extracted  from a  linear combination  of the  interferometric signal
 itself,  the beam  combiners have  no  dedicated photometric
 channels. This allows us to efficiently use all photons for 
 the interferometric combinations. 
 By design,  each interferometric  pair  simultaneously gives access to
 four  phase  states in  quadrature  (ABCD-like  but without  temporal
 modulation). These 4 measurements  allow the visibility amplitude and
 phase   to  be  retrieved   using  the   ABCD  method   described  in
 \cite{colavita}.   In practice, the  departure from  ideal quadrature
 forbids the  use of simple  algorithms and leads us  to consider a
 generalized algorithm capable of  handling a realistic description of
 the beam combiner properties.  

Throughout the paper, the outputs are identified with the index $m,l,k$,
such  as  $ml^{k}$,  where   $m,l$  are  the  interfering  beams,  and
$k$=[1..4],  the output for  this combination  (similarly, the
  A-C-B-D measurements of Fig.~\ref{fig:abcd}).  
The same nomenclature applies to functions.  In the case of Y-junctions, we denote
them  using the  index $m,l$  to specify  the beam  combination to
  which they are related, with $m$ corresponding to the actual beam that enters
  the Y-junction.  The index $k$ designates its two outputs.  For
  example, Y$_{12}^{1}$  and Y$_{12}^{2}$ are  the two outputs  of the
  Y-junction that  splits beam 1  in signals that will  interfere with
  beam 2.  Similarly, Y$_{21}^{1}$ and Y$_{21}^{2}$ are the outputs of
  the Y-junction  that splits  beam 2 into  signals that  will combine
  with  beam 1. We  use the  same notation  for the  couplers, \textit{e.g.}  C$_{24}^{1}$ and C$_{24}^{2}$ are  the outputs of
  the coupler corresponding to the combination of beam 2 and 4.  

With  this notation,  the intensity  recorded  at the  outputs of  the
combination of beams $m, l$ can be written~: 

\begin{equation}
\label{eq:interf}
  i_{ml}^{k}       =       N_{m}t_{ml}^{k}       +   N_{l}t_{lm}^{k}    +
  2V_{ml}^{obj}V_{ml}^{k}\sqrt{N_{m}t_{ml}^{k}N_{l}t_{lm}^{k}}
  cos(\varphi_{ml}^{k} + \varphi_{ml}^{p} +\varphi_{ml}^{obj}) 
\end{equation}

where  $N_{m}$  is  the  number  of  photons  in  the  $m$  beam and
$t_{ml}^{k}$ the  total transmission  of the $k$  output for  the $ml$
beam pair. $V_{ml}^{k}$ is the instrumental contrast, $\varphi_{ml}^{k}$ is the
instrumental phase introduced by the IO beam combiner between the
two interfering beams. $\varphi_{ml}^{p}$ is the residual atmospheric 
phase due to piston effects.  $V_{ml}^{obj}$ is the object visibility and $\varphi_{ml}^{obj}$ is its phase. 

\section{Laboratory set up}
In this section, we present the aims of the experiments, our testbed
as well as our operating mode for the data acquisition and processing.  

\subsection{Goals of the characterization}
With such beam combiners, all the information about the coherence of the
object is included in the way  the 4 pixels are related to each other,
\textit{including}  the instrumental contribution.   This contribution
has therefore to be known, \textit{i.e.}  fully calibrated.  

The relationship  between the  measured fluxes on  the pixels  and the
visibility amplitudes and phases of the object can be
expressed with a matrix representing the behavior of the instrument.  With an unresolved internal
source (\textit{i.e.}  $V_{ml}^{obj}= 1$ and $\varphi_{ml}^{obj}=
  0$) and without piston (\textit{i.e.} centered at zero OPD),
Eq.~(\ref{eq:interf}) becomes~: 

\begin{equation}
\label{eq:interf2}
  i_{ml}^{k}       =       N_{m}t_{ml}^{k}       +   N_{l}t_{lm}^{k}    +
  V_{ml}^{k}\sqrt{N_{m}N_{l}}X_{ml}^{k} 
\end{equation}
with  $X_{ml}^{k}    =\sqrt{t_{ml}^{k}t_{lm}^{k}}
cos(\varphi_{ml}^{k})$, a coefficient  different from 1, that corresponds
to the level at which the beam combiner conserves the coherence and that
depends on $\varphi_{ml}^{k}$, the  internal IO phase specific to the output $ml^{k}$.  
If one isolates    a  combination cell   \textit{[ml]}, the relation between the output intensity
and input number of photons (Eq.~(\ref{eq:interf2})) can be written as~:
\begin{equation}
\label{eq:interfjp}
\begin{tabular}{c}
$\left(
  \begin{array}{c}
    i_{ml}^{1}\\
    i_{ml}^{2}\\
    i_{ml}^{3}\\
    i_{ml}^{4}\\
  \end{array}
\right)
= 
\left(
  \begin{array}{ccc}
    t_{ml}^{1} & t_{lm}^{1} & V_{ml}^{1}X_{ml}^{1}(\varphi_{ml}^{1})\\
    t_{ml}^{2} & t_{lm}^{2} & V_{ml}^{2}X_{ml}^{2}(\varphi_{ml}^{2})\\
    t_{ml}^{3} & t_{lm}^{3} & V_{ml}^{3}X_{ml}^{3}(\varphi_{ml}^{3})\\
    t_{ml}^{4} & t_{lm}^{4} & V_{ml}^{4}X_{ml}^{4}(\varphi_{ml}^{4})\\
  \end{array}
\right) *
\left(
  \begin{array}{c}
    N_{m} \\
    N_{l} \\
    \sqrt{N_{m}N_{l}} \\ 
  \end{array}
\right)$
\end{tabular}
\end{equation}
$ml^{1}$ and  $ml^{2}$ correspond to  two outputs of the  same coupler
(the  same is valid  for outputs  $ml^{3}$ and  $ml^{4}$).  Therefore,
ideally, because energy is conserved at the output of a coupler the
following relations should apply: $\varphi_{ml}^{2} = \varphi_{ml}^{1}
+\pi$ and $\varphi_{ml}^{4}  = \varphi_{ml}^{3} +\pi$.  Similarly, the
beam combiner is ideally designed to introduce a phase quadrature 
between the outputs  therefore: $\varphi_{ml}^{3}  = \varphi_{ml}^{1}
+\pi/2$ and $\varphi_{ml}^{4} = \varphi_{ml}^{2} +\pi/2$. 

The overall behavior of the beam combiner can be generalized in a
matrix. When  considering all combinations, the matrix  should then be
constituted of similar blocks of a [$4\times3$] matrix with zero elsewhere.  In
reality, crossing terms appear both as incoherent and 
coherent contributions,  and the actual  outgoing intensities
should be described using a general matrix of $24\times10$~terms~:

\begin{center}
\begin{equation}
\label{eq:bigmat}
\left(
 \begin{array}{cccccccccc}
 t_{12}^{1} & t_{21}^{1} &  &   & V_{12}^{1}X_{12}^{1} &  &  &
  &  & \\
 : & & & : & & & :& & & :\\
 : & & & : & & & : & & & :\\
  &  & t_{34}^{4} & t_{43}^{4} &  &  &  &  &  & V_{34}^{4}X_{34}^{4}\\
 \end{array}
 \right)*
 \left(
 \begin{array}{c}
 N_{1} \\
 :\\
 N_{4} \\
 \sqrt{N_{1}N_{2}} \\ 
 :\\
 \sqrt{N_{3}N_{4}} \\ 
 \end{array}
 \right) 
\end{equation}
\end{center}

\bigskip
This matrix, called the V2PM (visibility to pixel matrix), in accordance to previous work on
multiaxial interferometers \citep{tatullilebouquin06}, completely characterizes the instrumental
behavior of the beam combiner \textit{e.g.}  the transmission, the visibility,
the phase relations and the parasite flux.  The ultimate goal of such a study
would be to precisely estimate and calibrate it \citep{lacour2008}. However, it 
is out of the scope of such a paper to present and discuss a full characterization of
the global beam combiner  matrix. We prefer to focus  on characterizing the
individual    tricoupler   functions,   Y-junctions,    couplers   and
phase-shifting devices  described by Eq.~(\ref{eq:interfjp}) as
well  as  the  global   routing  of  the  incoherent  flux  (so-called
crosstalk)  inside the  beam  combiner. As  it  will be  seen
later, the  crosstalk terms (related to the  crossing terms in
  Eq.~(\ref{eq:bigmat}))  are  sufficiently  small, which  justifies  this
approach.  

To  reach this goal,  we set  up a  laboratory testbed  and tested  the IO
beam combiners  through  photometric  and interferometric  measurements  to
calibrate this instrumental matrix.  
The following quantities, as well as their dependence on wavelength,
have been measured:
\begin{enumerate}
\item the so called ``normalized kappa matrix'' $\kappa^{k}_{ml} =\frac{t^{k}_{ml}}{\sum_{m}\sum_{l \not m} t^{k}_{ml}} $;
\item instrumental contrast $V^{k}_{ml}$;
\item      instrumental     phase      shift      between     outputs,
  $\varphi_{ml}^{k}$-$\varphi_{ml}^{k+2}$,   with   $\varphi_{ml}^{k}$
  being the individual phase of the output signal $ml^{k}$; 
\item instrumental  closure  phase $\Phi_{mlj}^{k}  =  \varphi_{ml}^{k}+
  \varphi_{lj}^{k} + \varphi_{jm}^{k}$;
\end{enumerate}

\subsection{Testbed description}

\begin{figure}
  \begin{center}
    \includegraphics[width=0.5\textwidth]{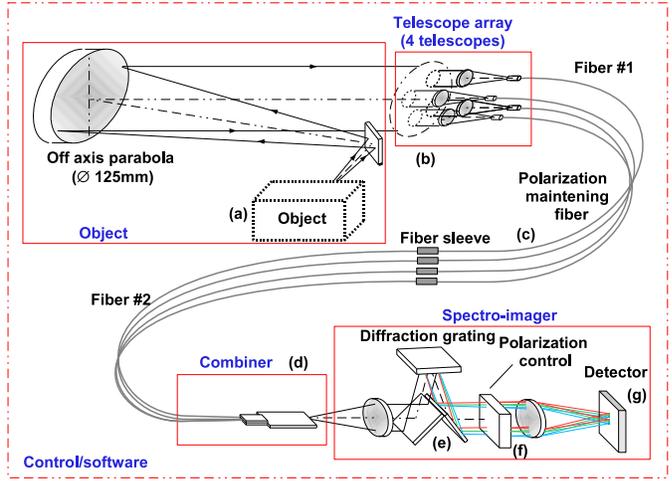}
    \caption{\label{fig:banc}   Schematic  view   of   the  laboratory
      testbed simulating the VLTI (see text for details).} 
  \end{center}
\end{figure}

We designed a dedicated interferometric testbed capable of simulating
an  8 telescope interferometer  \citep{jocou07}. Figure~\ref{fig:banc}
describes individual functions of the setup.  
The bench includes various items~:
\begin{itemize}
\item[] (a) an object simulator that  can reproduce a single star or a
    binary star with an adjustable flux ratio
\item[] (b) up to 8 optical devices simulating telescopes and coupling the
  light into single-mode polarization-maintaining fibers 
\item[] (c)  optical path  compensation and modulation  devices (delay
  lines of a few mm long)
\item[] (d) an IO beam combiner 
\item[] (e) a spectrometer 
\item[] (f) a Wollaston prism to split the linear polarizations
\item[] (g) an infrared detector. 
\end{itemize}

\begin{figure}[t]
  \begin{center}
    \includegraphics[width=0.35\textwidth]{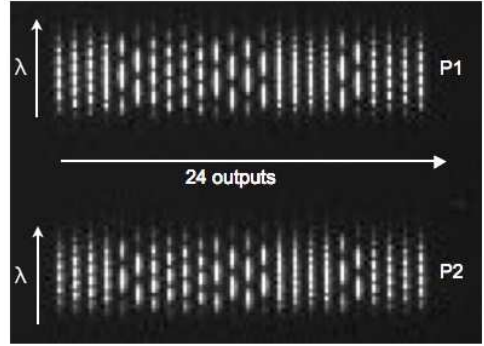}
    \caption{\label{fig:dispersion}  A   detector  image  of   the  24
      outputs, obtained  when using  the  spectrograph and  a
      Wollaston prism  (splitting the two linear  polarizations P1 and
      P2).  The patterns are due to various non zero OPD for the
      different beam combinations.} 
\end{center}
\end{figure}

All laboratory tests were carried out in the H band with light
sources of different coherence lengths.  The object
simulator can reproduce a single star as well as a binary star. In the
latter case, its design is based on an optical setup that mimics a Michelson
interferometer, but  with a tilted mirror  in one of its arm and with
an unbalanced pathlength between the two arms.  It produces two
non-coherent luminous spots 
simulating a binary star, whose separation can be adjusted by tilting 
the mirrors.  This setup will be used to characterize the dynamics of
the testbed. The image is placed at the focal plane of a F/5
collimator to produce a $100\,\mbox{mm}$ diameter collimated beam. The
wavefront is sampled by up to 8 telescopes that can be set to
reproduce a replica  of the VLTI entrance pupil.  These telescopes are
made up of an F/5 gradium lens feeding a polarization 
maintaining  fiber  and  are  placed  in  the  collimated  beam.  Each
telescope is mounted on an individual module with tip-tilt adjustments
and a motorized translation stage  (delay lines).  
Shutters are used to block the  light of each
telescope. The fibers, that  have been equalized to limit differential
effects ($\Delta L$=1mm), are gathered in a V-groove 
chip that feeds the IO beam combiner.  A diffracting grating provides 15 spectral
channels through the H band, while the Wollaston prism splits the linear polarization states to improve the instrumental transfer
function.  These  two last  elements, that can  be placed  and removed
  easily depending on the need, are located in an afocal mount with a 
magnification of 1. The detector is a near-infrared InGaAs PICNIC chip, with
40$\mu$m-large pixels. In the case of wide band measurements, each
waveguide      output      is      imaged      on      one      single
pixel.  Figure~\ref{fig:dispersion} is a detector image obtained when both the
spectrograph  and  the Wollaston  prism  are  used.  It shows  the  24
beam combiner outputs spectrally dispersed along the vertical 
direction.   P1  and P2  correspond  to  the  two linear  polarization
states.  

\begin{table}[t]
\begin{center}
\begin{tabular}{c} 
\begin{tabular}{|cccccc|}
\hline
Step & Sh1 & Sh2 & Sh3 & Sh4 & Measurement\\
\hline
1 & X & X & X & X & B$_{\rm g}$\\
2 & O & X & X & X & $P_{1}$\\
3 & X & O & X & X & $P_{2}$\\
4 & X & X & O & X & $P_{3}$\\
5 & X & X & X & O & $P_{4}$\\
6 & O & O & O & O & $I$\\
\hline
\end{tabular}
\\
\\
\begin{tabular}{c}
Sh = shutter; O=open; X=closed
\\
\end{tabular}
\end{tabular}
\caption{\label{tab:calib} Experimental protocol including
  photometric and  interferometric measurements.} 
\end{center}
\end{table}

\subsection{Data acquisition and processing}
\subsubsection{Protocol}
The sequence  of acquisitions performed to  characterize the beam combiners
consists of 6 steps that  can be done with or without spectral
  dispersion (see Table~\ref{tab:calib}).  
Step~1 is a background measurement with
all shutters closed to prevent any light from propagating through the
instrument.  The 4 consecutive steps are measurements with only one 
beam at the time that give access to the flux splitting ratios in the
couplers  and  tricouplers.  Finally,  Step~6 is  the  interferometric
combination  of all input  beams. All  measurements are  repeated 1024
times.  

The first 5 steps are used to validate the design in terms of
photometry   (light  routing,   transmission  and   splitting  ratios,
undesired  flux, together with  their wavelength  dependence).  Step~6
leads to the determination of the value, stability and chromaticity 
of instrumental contrasts  and closure phases as well  as of the phase
relations between phase-shifted  outputs, supposedly in quadrature. To
get  a complete  and independent  laboratory characterization  of each
output of the tested beam combiners, the interferometric
measurements presented  in this  paper are obtained  with OPD
modulation and polarization splitting on a point-like source.

\begin{figure*}
  \begin{center}
\begin{tabular}{cc}
  \includegraphics[width=0.4\textwidth]{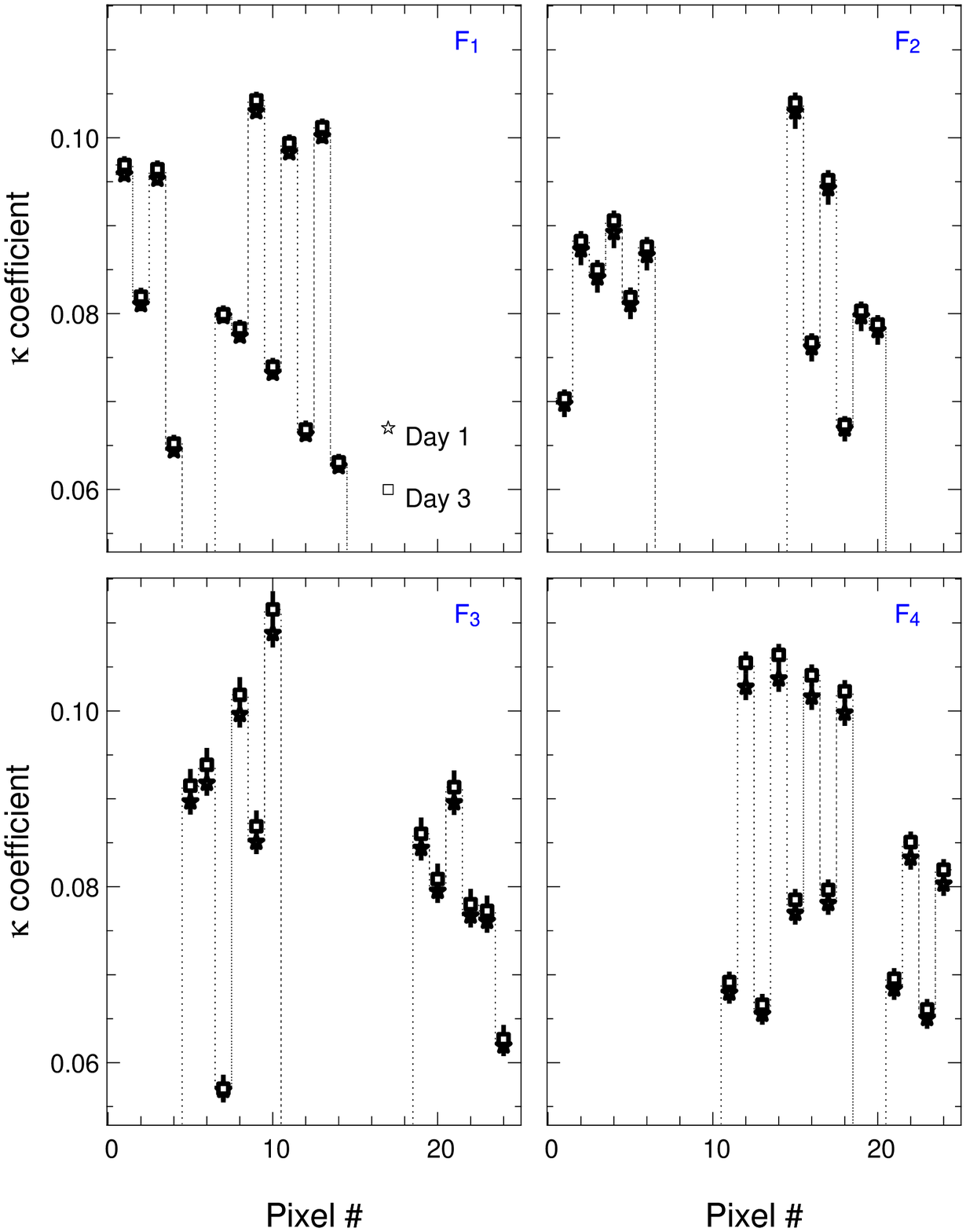}
& 
     \includegraphics[width=0.39\textwidth]{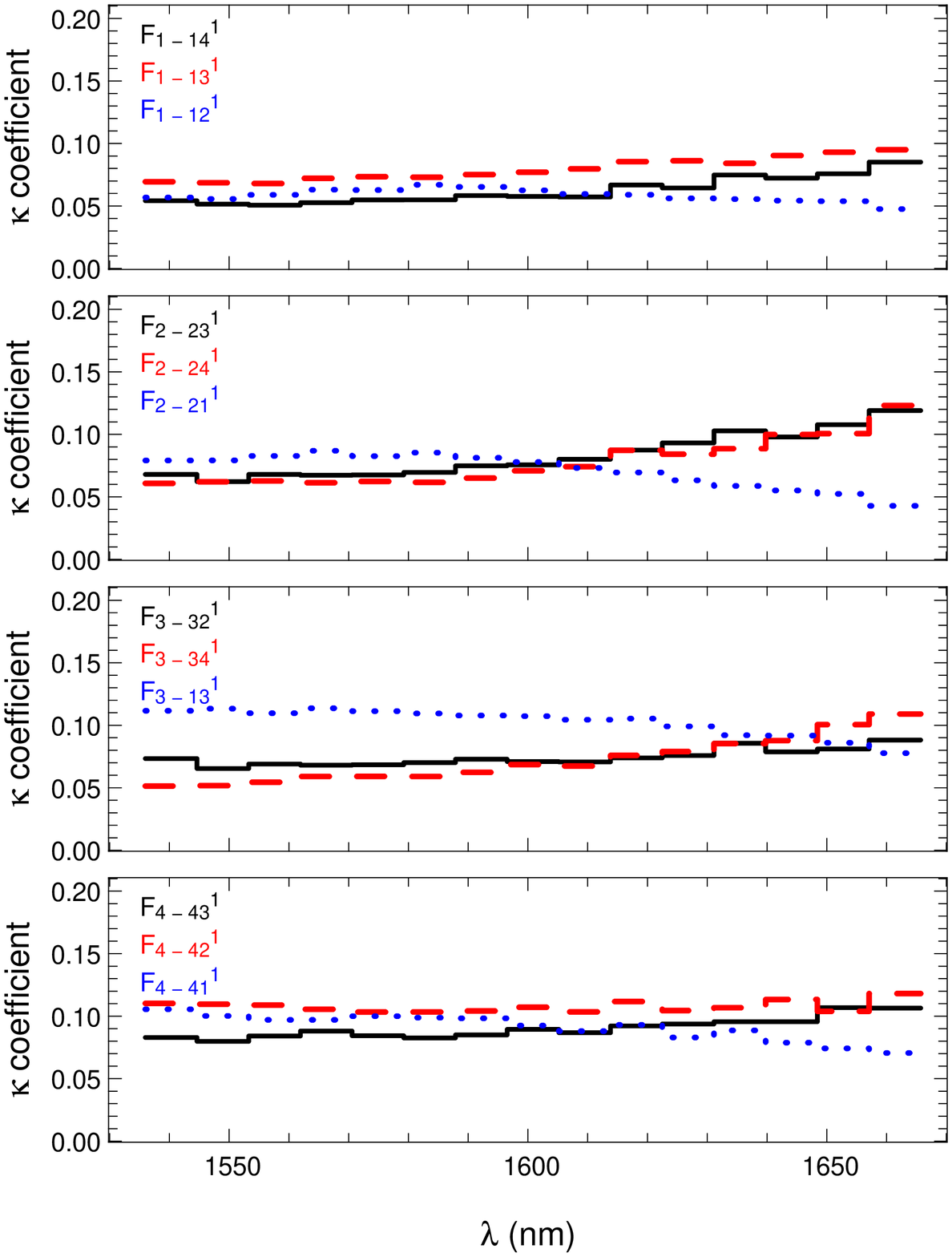}
\end{tabular}
\caption{\label{fig:photocoeff} Left: kappa matrix photometric coefficients obtained when
  the  light is injected  in one  input at  a time  (F$_{1..4}$) for
  Chip1.  The stars and squares show results obtained on two different
  days of experiments (called Day~1 and Day~3). Pixels from 1 to
  24 are  the 24  outputs of the  beam combiner, identified as  $12^{1}$ to
  $34^{4}$ in Table \ref{tab:coeffphotom4tabcd}. Right: variation
  of the  kappa matrix  photometric coefficients with  wavelength over
  the H~band range, measured on Chip2. The four
  panels correspond  to the light injection  in one input  at the time
  (\textit{e.g.} F$_{1}$ for the injection in input 1).
  Inside each of them, among the 12 outputs illuminated, only 3 are
  plotted (in full, dashed, and dotted lines), corresponding to each combination cell.} 
\end{center}
\end{figure*}

\subsubsection{Data reduction}
The  data  processing  derives  four  quantities~:  the  kappa  matrix
(\textit{i.e.}  the  photometric  contribution  of each  beam  to  the
interferogram); the instrumental  contrast; the phase-shift induced by
the devices and the closure phases. 

The kappa matrix  is extracted from each individual  set of data where
only one input is illuminated (Table~\ref{tab:calib}, Steps 2-5). 
\\
\indent
The  instrumental contrast  is computed  using a  classical visibility
estimator on the interferograms to evaluate the 
coherence.   This consists  of estimating  the envelope  amplitude and
calibrating  for  the photometric  inbalance  between the  interfering
beams.   Since  our experimental  data  are  obtained  with a  SNR  of
$\sim$100, the use of such a simple estimator is appropriate. 
\\
\indent
The phase-shift  between two outputs is computed  as the phase
  of the complex  product of the Fourier spectra  corresponding to the
  two  signals.  For  two phase-shifted  outputs related  to the
  combination   cell  [ml]  (written   \textit{$ml^{1}$}  and
  \textit{$ml^{3}$}, \textit{i.e.} outputs $k=1$ 
  and   $k=3$   respectively),    it   is   calculated   as   follows:
  $$\varphi_{ml}^{1}-\varphi_{ml}^{3}                                 =
  \textrm{atan}\left(\frac{\Im[F_{ml^{1}}(\nu_{ml,1})F_{ml^{3}}^{*}(\nu_{ml,3})]}{\Re[F_{ml^{1}}(\nu_{ml,1})F_{ml^{3}}^{*}(\nu_{ml,3})]}\right)$$ 
  where $\Im$,$\Re$ stand for the imaginary and real parts respectively. 
  F$_{ml}^{k}(\nu_{ml,k})$ (here with $k=1,3$) is the intensity of the
  Fourier  spectrum  of the  signal  \textit{$ml^{k}$},  taken at  the
  maximum value (to which $\nu_{ml,k}$ corresponds).
\\
\indent
The closure phase is measured using a triplet of telescopes
\textit{i.e.}  from three pairwise combinations.  It is calculated as
  the  phase of  the bispectrum,  that is  the complex  product  of the
  corresponding  three Fourier spectra.   Since each  beam combination
  produces 4  phase-shifted output signals, there  are 4 closure-phase
  signals per telescope triangle.  
  For the telescope triangle [mln], the closure-phase derived from 
  the output $k$ can be written as: 
$$\Phi_{mlj}^{k}                                                      =
\textrm{atan}\left(\frac{\Im[F_{ml}^{k}(\nu_{ml,k})F_{lj}^{k}(\nu_{lj,k})F_{mj}^{k*}(\nu_{mj,k})]}{\Re[F_{ml}^{k}(\nu_{ml,k})F_{lj}^{k}(\nu_{lj,k})F_{mj}^{k*}(\nu_{mj,k})}\right)$$ 

The closure phase is computed with the constraint that the frequencies
respect   the   closure   relation   (\textit{e.g.}   $\nu_{mj,k}$   =
$\nu_{ml,k}$ + $\nu_{lj,k}$). Definitions of F$_{ml}^{k}(\nu_{ml,k})$,
F$_{lj}^{k}(\nu_{lj,k})$ and F$_{mj}^{k}(\nu_{mj,k})$ are identical to
the phase-shift case.

\bigskip

The methodology for data reduction with spectral dispersion is
  identical.  From  all illuminated  pixels of the  detector (in  a case
  such  as in  Fig.~\ref{fig:dispersion}),  we measure  interferograms
  from  which  we  derive   the  chromatic  behavior  of  instrumental
  quantities.

\section{Results}
In this section, we present  the results of the laboratory experiments
obtained on a point-like source, in terms of flux throughput 
and routing, instrumental contrasts, phase-shifts, and closure phases.  
The results correspond to  two different IO chips manufactured
  in the same wafer (called Chip1 and Chip2 in all tables) in broad band, as well as with
spectral dispersion for the second chip only.  

 For the latter experiment, we report in all tables the average values as well as the
\textit{amplitude}  of   the  variation  over   the  wavelength  range
(\textit{i.e.}  $|X_{max,\lambda}-X_{min,\lambda}|$, for the 
instrumental quantity  $X$).  We  refer  to  the  latter as  \textit{chromaticity}  in  the
text. 
Detailed studies that relate the performance to the IO design and simulations will be given in a following paper 
(Labeye  et  al.~2009,~in~prep.).  All  results  are  commented on  in
Sect.~\ref{sec:dis}.  The notations used in the tables and figures 
are the same as defined in Sect.~\ref{sec:tech}.

\subsection{Photometric measurements}
\textit{Transmission:} 
When injecting 100 photons in one input, the transmission is the total
number of photons detected at all outputs.
The  overall transmission budget includes  the coupling
  efficiency from the telescope point spread function to the fiber, as
  well as  the propagation  losses inside the  fibers and the  IO chip
  transmission.  
  The latter quantity is determined in \textit{broad band} by using a fiber at each input and
  at  each  output,   and  the  measured  flux  is   normalized  by  a
  fiber-to-fiber  transmission.  The measurement  gives about  65\% in
  the  H~band  for  the  transmission  of  the  IO  chip  itself.  The
  telescope-to-fiber  coupling  is  ideally  $\approx 80\%$  using  a
  perfect  circular pupil without  central obscuration.   The silicate
  fiber  transmission in  the  H band  is  excellent ($\leq  5dB/km$),
  consequently,  for the  2m fiber  lengths  that we  are using,  the
  corresponding transmission is $\approx 99\%$.  The total throughput of
  the 'fibers+combiner' is therefore $\approx 64\%$.

\begin{table*}
 \begin{center}
 \caption{\label{tab:coeffphotom4tabcd} Kappa matrix photometric    coefficients
   obtained  when  the  light is injected  only  in  the  4th  input
   (F$_{4}$).  The  first  two  lines  correspond  to  the  wide  band
   experiments  while   the  third  gives  the   average  value  over
   wavelength, as  well as the  peak-to-valley amplitude over
   the wavelength range (chromaticity).  Values are divided by $10^{4}$.  Bold
   numbers, preceded by a star, indicate the illuminated outputs.  }  

 \begin{tabular}{c}
 \begin{tabular}{c||c c c c c c c c c c c c c}
\hline
\hline
 Output &  $12^{1}$ & $12^{2}$ & $12^{3}$  & $12^{4}$ &
 $23^{1}$   &  $23^{2}$   & $13^{1}$   &   $13^{2}$  &
 $13^{3}$   &   $13^{4}$   &  \textbf{\large{$*14^{1}$}}   &
 \textbf{\large{$*14^{2}$}} &   \textbf{\large{$*14^{3}$}}\\
 \hline
 Chip 1  & 0.2$\pm$9 &  0.2$\pm$9 & 0.8$\pm$9  & 0.8$\pm$9 &  1$\pm$9 &
 0.8$\pm$  9&  1$\pm$9 &  1$\pm$9&  3$\pm$9  &  8$\pm$9 &  706$\pm$9  &
 1069$\pm$9 & 681$\pm$10 \\

 Chip 2  & 0.5$\pm$3 & 0.5$\pm$3  & 0.8$\pm$ 3& 0.5$\pm$3  & 2$\pm$3 &
 3$\pm$3  &  1$\pm$ 3&  1$\pm$  3& 2$\pm$  3&  3$\pm$  3& 801$\pm$  4&
 932$\pm$ 4&751$\pm$3 \\

avg / $\Delta \lambda$ & 1/2 & 2/6 & 1/4 & 2/6 & 3/9 & 4/6 & 4/5 & 3/5
& 5/5 & 6/11 & 621/344 & 926/357 & 624/237\\ 
 \end{tabular}
 \\
 \\
 \begin{tabular}{ c||c c c c c c c c c c c }
\hline
\hline
 Output  &  \textbf{\large{$*14^{4}$}} &  \textbf{\large{$*24^{1}$}}  &  \textbf{\large{$*24^{2}$}} &
  \textbf{\large{$*24^{3}$}}  & \textbf{\large{$*24^{4}$}} &  $23^{3}$ &
  $23^{4}$ & \textbf{  \large{$*34^{1}$}} & \textbf{\large{$*34^{2}$}} &
  \textbf{ \large{$*34^{3}$}} & \textbf{\large{$*34^{4}$}} \\ 
 \hline
 Chip 1  &  1078$\pm$10 & 800$\pm$10 & 1055$\pm$10 & 811$\pm$ 10&
 1037$\pm$ 9& 4$\pm$10 & 10$\pm$9 & 710$\pm$9 & 865$\pm$9 & 675$\pm$ 9&
 834$\pm$ 9  \\

 Chip 2 &  944$\pm$4 & 784$\pm$4 & 930$\pm$4& 691$\pm$4 & 990$\pm$4 &4$\pm$3 & 3 $\pm$3& 693$\pm$3 & 896$\pm$4 & 691$\pm$4 & 874$\pm$4\\

avg /  $\Delta \lambda$  & 953/184  & 794/270 &  1032/150 &  820/171 &
1055/124 & 8/6 & 6/8 & 585/194 & 865/412 & 766/360 & 913/573 \\

 \end{tabular}
 \end{tabular}
 \end{center}
 \end{table*}

 \begin{figure*}
   \begin{center}
     \begin{tabular}{ccc}
       \includegraphics[width=0.32\textwidth]{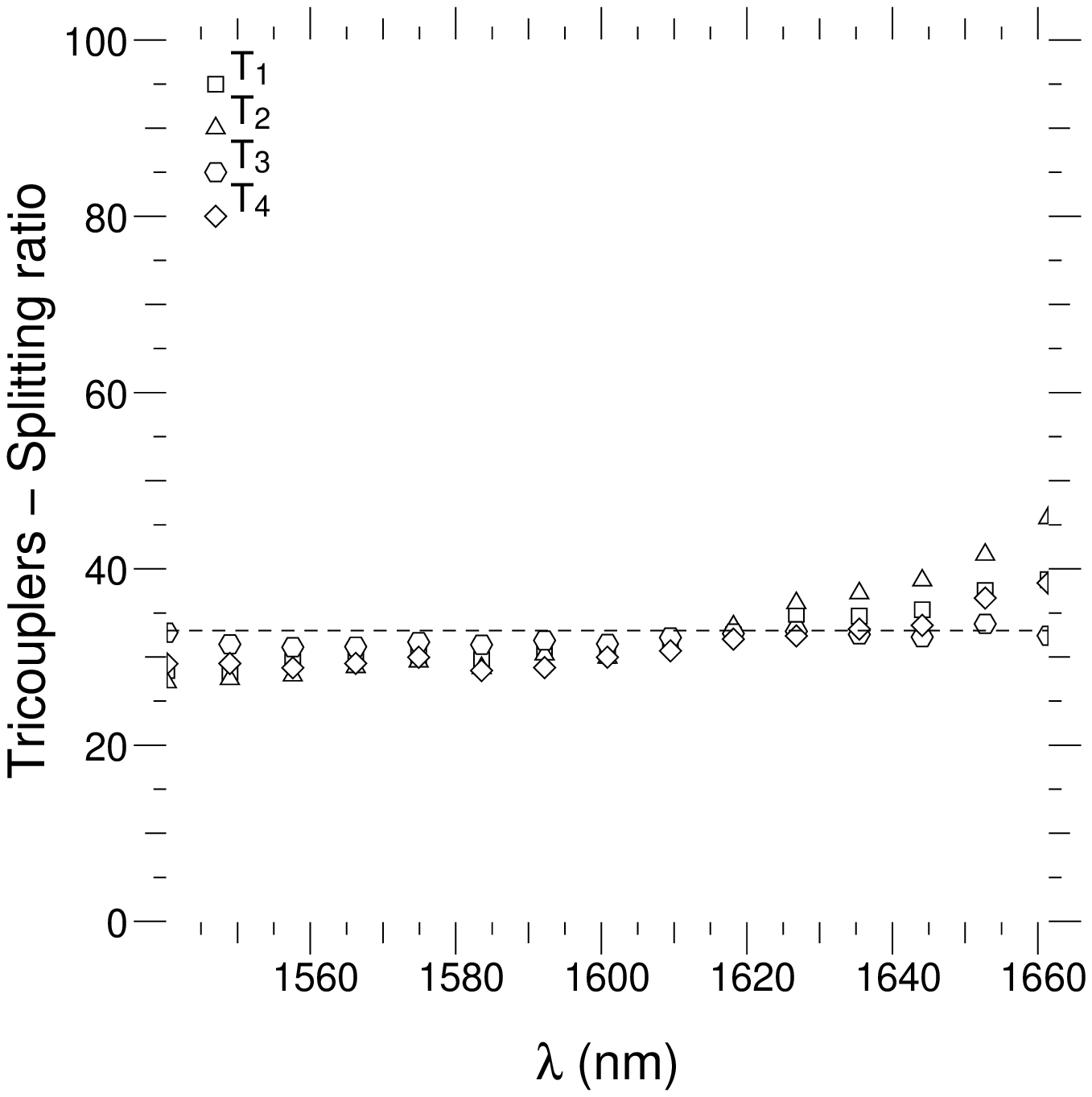}
       & 
       \includegraphics[width=0.32\textwidth]{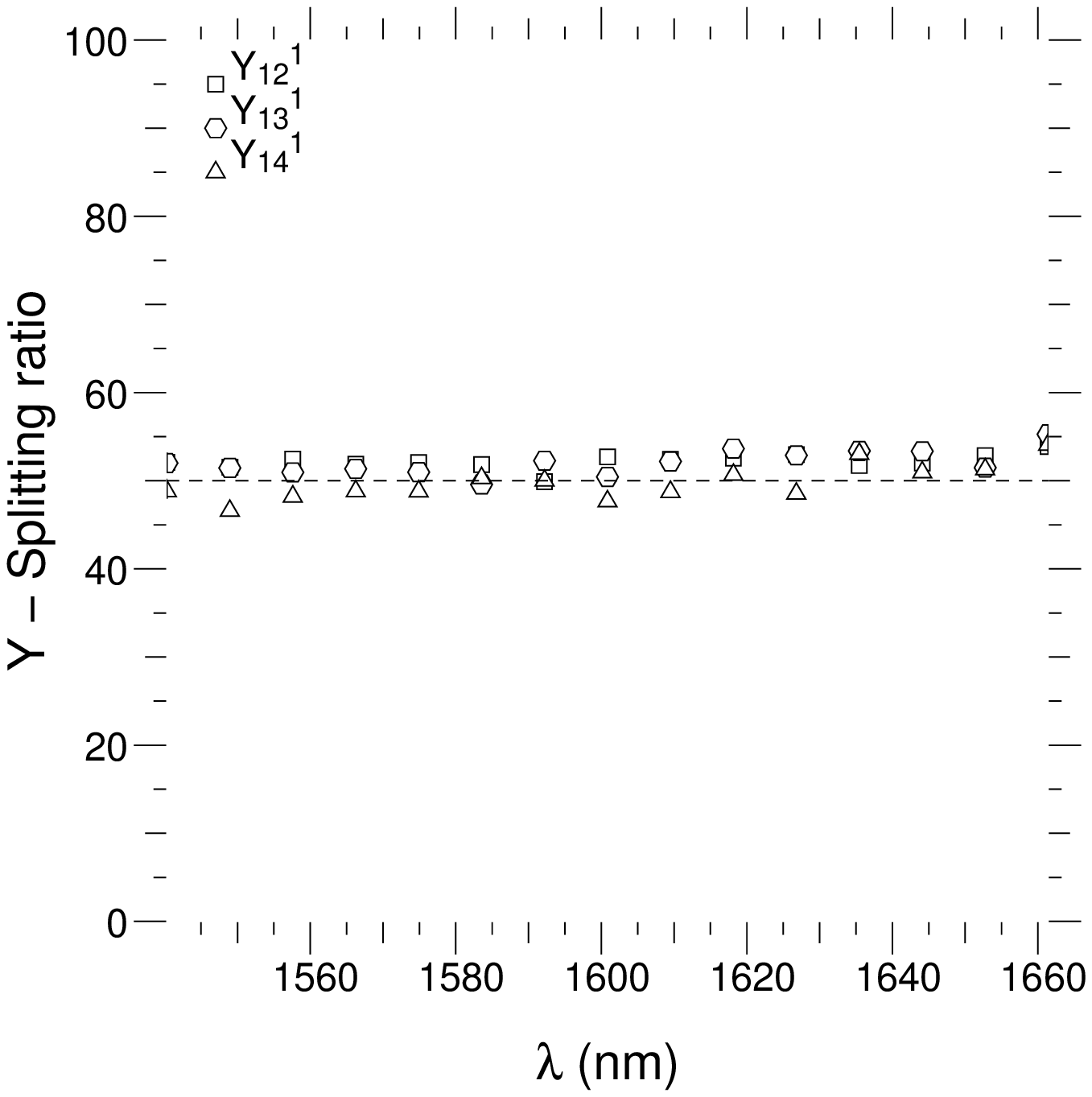}
      & 
       \includegraphics[width=0.32\textwidth]{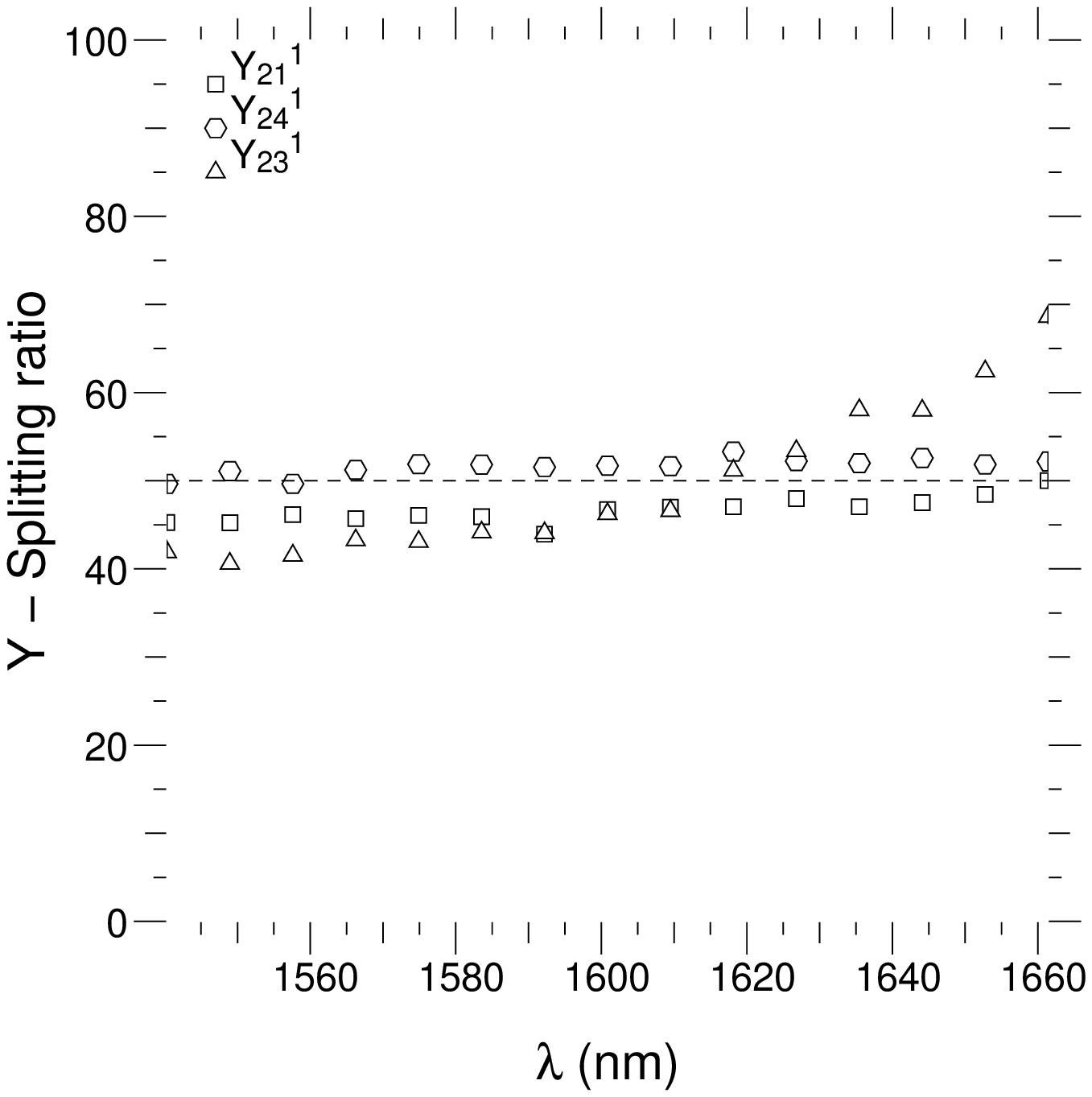}
     \end{tabular}
     \caption{\label{fig:y_lam} Variation with wavelength of the
       tricoupler  splitting  ratio  (left)  and of  the  Y-junctions
       (middle, right).  Middle: three Y-junction
       splitting ratios  corresponding to the illumination  in the first
       input (these Y-junctions are related to the [12], [13]
       and [14]  combinations); Right: three  Y-junctions splitting
       ratios obtained when injecting the light in the second input. 
       For clarity, only one flux ratio coefficient (among 2 or 3) is
       plotted  for each  function.  The error for  each
       spectral  channel is  estimated from  the dispersion  over 1024
       measurements and is smaller than the symbol sizes.  The
       theoretical values of 33\% for  the tricoupler, and 50\% for the
       Y-junctions, are  given by the horizontal  dashed line.  Values
       for all  functions are given  in Table~\ref{tab:tricouplers_wb}
       and \ref{tab:jY}.}  
 \end{center}
 \end{figure*}

\bigskip
\textit{Flux routing and individual IO functions:}
For   the   following   paragraphs,    we   use   Step~1   to   Step~5
(Table~\ref{tab:calib}).  Figure~\ref{fig:photocoeff}, left, gives the 
broad band photometric coefficients for the 24 beam combiner outputs obtained when
shutters prevent three telescope beams from propagating through the
chip. These  coefficients are  defined, for each  pixel, by  the ratio
between the flux detected on one pixel  and the sum of the flux on all
outputs.  As  expected  in such  a  case,  12  out  of 24  pixels  are
illuminated. Two measurements taken on 2 different days are
compared  (stars  and  squared  symbols)  showing  very  small  time
  variability. Table~\ref{tab:coeffphotom4tabcd} gives an example
of the averaged photometric  coefficients for all 24 outputs, obtained
when the light is injected in the \textit{4th} input. 

With  the same experiments  using spectral  dispersion, one  can derive
their dependence on wavelength. Figure~\ref{fig:photocoeff}, right, gives their
variation   with    wavelength     over     the    spectral
range.  Table~\ref{tab:coeffphotom4tabcd}  also gives the chromaticity
of   photometric   coefficients  obtained   when   injecting  in   the
\textit{4th} input.\\ 
To be more general, Table~\ref{tab:relat_kappa} gives measurements in all cases, \textit{i.e.} when the light is injected
in  all  4  inputs,  one  at  a  time.  We  only  provide  the  values
corresponding to the outputs presenting the minimum and maximum chromaticity as well
as the average  chromaticity over the 12 signals.   Because there is
an important spread across  the photometric coefficients values (see
Table~\ref{tab:coeffphotom4tabcd}),  the chromaticity is  given with
respect to the coefficient value obtained when averaging over the spectral band
(\textit{i.e.} divided by this value).

\begin{table}[t]
 \begin{center}
 \caption{\label{tab:relat_kappa} Minimum, maximum  and average of the
   photometric  coefficients  chromaticity  among the  12  illuminated
   outputs from each injection.} 
 \begin{tabular}{ c||c c c }
\hline
\hline
Injection in input \#& Average  & Minimum  & Maximum  \\
\hline
1 & 30\% & 14\% & 48\% \\
2 & 51\% & 30\% & 81\% \\
3 & 58\% & 27\% & 80\% \\ 
4 & 35\% & 12\% & 63\% \\ 
 \end{tabular}
 \end{center}
 \end{table}

 \begin{table*}
 \begin{center}
 \caption{\label{tab:tricouplers_wb}   Tricoupler   splitting   ratio
   measured in wide band.  The dispersion (rms) over 1024 measurements is
   0.1\%. The  two first lines  correspond to the wide  band experiments
   while the third gives the  average value and the variation amplitude
   over the wavelength range.} 
 \begin{tabular}{c||cccc}
 \hline
 \hline
 T & T$_{1}$ & T$_{2}$ & T$_{3}$ & T$_{4}$\\
 \hline
 Chip 1 & 33.7 - 33.3 - 32.8 &
 33.2 - 32.6 - 34.1 & 34.5 - 35.0 - 30.3 & 34.2 - 35.9 - 29.8 \\

 Chip 2 & 31.9 - 35.3 - 32.7 &
 33.8 - 34.0 - 32.1 & 32.4 - 36.4 - 31.1 & 34.3 - 34.0 - 31.6 \\

avg /$\Delta \lambda$ & 32.2/10.5 - 37.2/2.9 - 30.5/11.9 & 
 32.9/18.6 - 31.4/6.2 - 35.6/13.2 & 32.1/2.6 - 37.0/17.9 - 30.8/19.0 &
 31.3/9.9 - 37.1/5.0 - 31.4/14.2\\
 \end{tabular}
 \end{center}
 \end{table*}

\begin{table}[t]
 \begin{center}
 \caption{\label{tab:relat_tri}    Minimum,   maximum    and   average
   chromaticity  of  tricoupler  splitting  ratio  (among  the  three
   outputs   of   each  tricoupler).    T$_{1}$   is  the   tricoupler
   corresponding to input 1.} 
 \begin{tabular}{ c||c c c }
\hline
\hline
Tricoupler & Average & Minimum & Maximum \\
\hline
T$_{1}$ & 8\% & 3\% & 12\% \\
T$_{2}$ & 13\% & 6\% & 19\% \\
T$_{3}$ & 13\% & 3\% & 19\% \\ 
T$_{4}$ & 9\% & 5\% & 14\% \\ 
\end{tabular}
\end{center}
\end{table}


From  these coefficients,  in  both wide  band and  spectrally
  dispersed experiments, we determine the splitting ratio of the
different optical  functions (tricouplers, Y-junctions  and couplers),
under  the assumption that  all functions  are ideal  (\textit{i.e} no
photon loss;  $\sum_{i} x_{i}=100\%$,  with $x_{i}$ a  splitting ratio
coefficient).  For  the  sake  of  clarity, for  each  Y-junction  and
coupler, only one value out of the two splitting ratio coefficients is
given  in  the  tables,  since  the second  output  is  obviously  its
complementary to 100\%.

\vskip 0.09cm
\textit{Tricouplers:} 
Table~\ref{tab:tricouplers_wb} gives such values for the 4 tricoupler
showing flux splitting ratios close to 33\% for both chips, in wide
band.  The best  flux  separation is  33.7/33.4/32.8,  for the  three
outputs, with  a rms over 1024 measurements  of 0.1\%.  These
values are similar a few days later with a variation from 0.1\%
to 1.5\%.  \\
The results are comparable with the spectral dispersion. In this case,
the  closest  splitting  ratio   from  33\%  is  32.9/31.4/35.6  $\pm$
0.1\%. Figure~\ref{fig:y_lam}, left, presents the variation of the 
tricoupler splitting ratio with wavelength.  Table~\ref{tab:relat_tri}
gives   the  minimum,   maximum  and   average  chromaticity   of  the
tricoupler splitting ratio, among the 3 outputs of each of the four
tricouplers of Chip2.  


 \begin{table*}
 \begin{center}
 \caption{\label{tab:jY} Y-junction splitting ratio. The dispersion (rms) over 1024
   measurements is 0.1\%. The two first lines correspond to the wide
 band experiments  while the  third gives the  average value  and the
 variation amplitude over the wavelength range. }
 \begin{tabular}{ c||c c  c c  c c  c c  c c  c c }
\hline
\hline
 Y&  Y$_{12}^{1}$  &  Y$_{12}^{2}$  &  Y$_{13}^{1}$  &  Y$_{13}^{2}$  &
 Y$_{14}^{1}$ & Y$_{14}^{2}$& Y$_{24}^{1}$ & Y$_{24}^{2}$ & Y$_{23}^{1}$ & Y$_{23}^{2}$ & Y$_{34}^{1}$ & Y$_{34}^{2}$\\
 \hline
 Chip 1 & 52.5 & 47.4 & 47.0 & 44.4 & 50.2 & 50.2 & 52.6 & 50.0 &
 51.5 & 52.6  & 54.7  & 51.0  \\

 Chip 2 & 48.7 & 45.7 & 52.2 & 49.8 & 51.2 & 50.5 & 54.1 & 50.4 & 53.0 & 53.2 & 52.2 & 50.3 \\

 avg / $\Delta \lambda$ & 52.1/4.0  & 46.6/6.0 & 52.0/5.7 & 48.8/8.0 &
 49.7/7.5 & 49.3/7.4  & 51.6/3.7 & 49.3/4.2 &  49.5/28.0 & 50.1/25.9 &
 49.5/4.5 & 46.5/6.2 \\ 
 \end{tabular}

 \end{center}
 \end{table*}

\bigskip
\textit{Y-junctions:} 
Table~\ref{tab:jY} gives the splitting ratio for the 12
Y-junctions. The values, measured in broad band, are close to
50\%, with  a 0.1\%-dispersion over 1024 points.  Variations are small
from one day to another (2.9\% maximum). \\
With spectral dispersion, the splitting ratio are similar
to the  broad band measurements.   Figure~\ref{fig:y_lam}, middle and
right, gives their
wavelength-dependence.  For clarity,  only extreme behaviors are shown
in the  figure, with  the smallest (middle  plot) and  greatest (right
plot) variations over the H band.   Out of the 12 Y-junctions, 10 show
a maximum variation inferior to 9\% over the spanned range of the H band,
while  2 show  a strong  variation  of about  26 and  28\%. These  two
Y-junctions are  the closest to  the inputs (combination  [23]).  Over
all the Y-junctions, the average chromaticity is 9.2\%.

 \begin{figure*}[t]
   \begin{center}
     \begin{tabular}{cc}
       \includegraphics[width=0.35\textwidth]{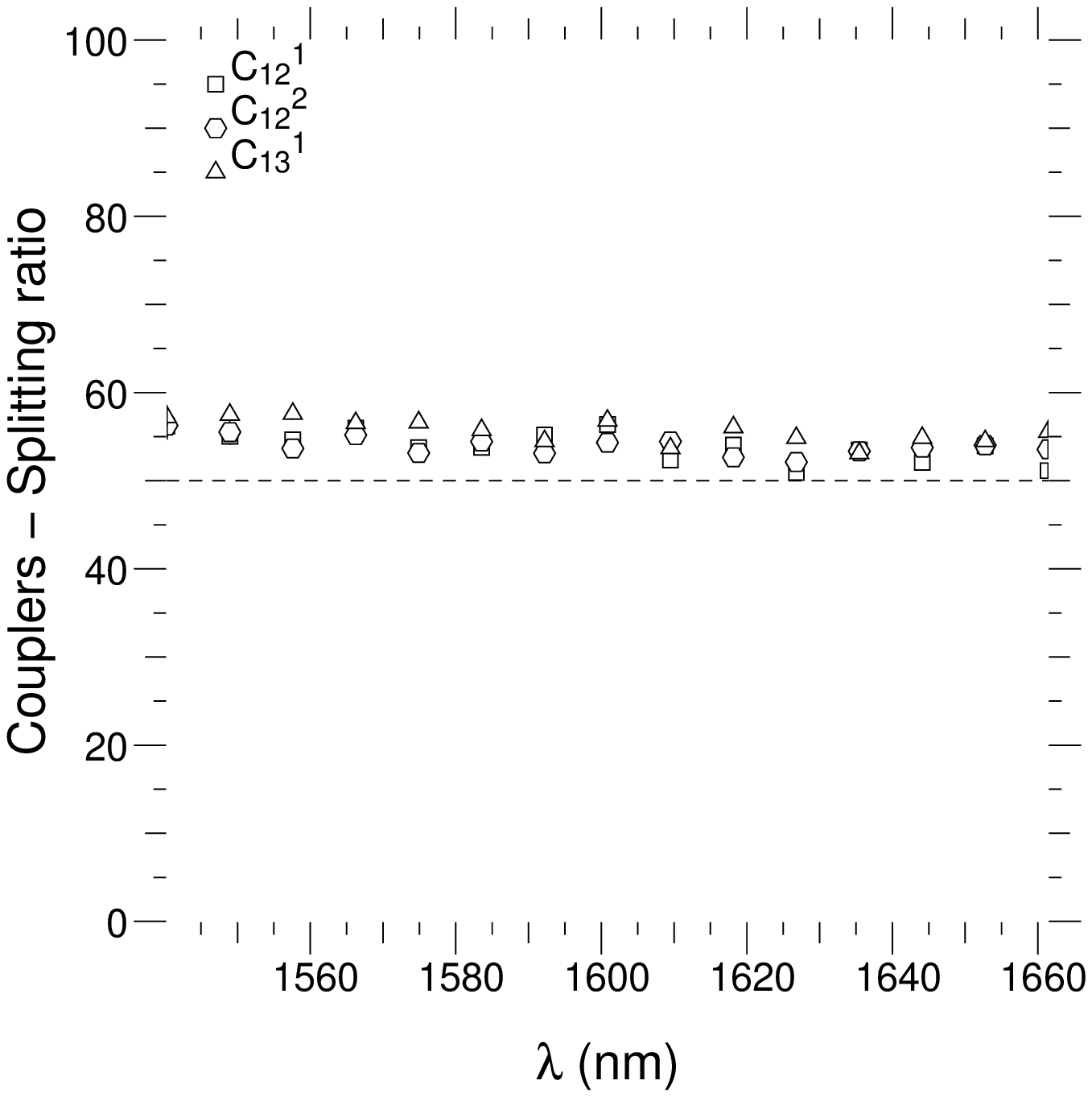}
       & 
       \includegraphics[width=0.35\textwidth]{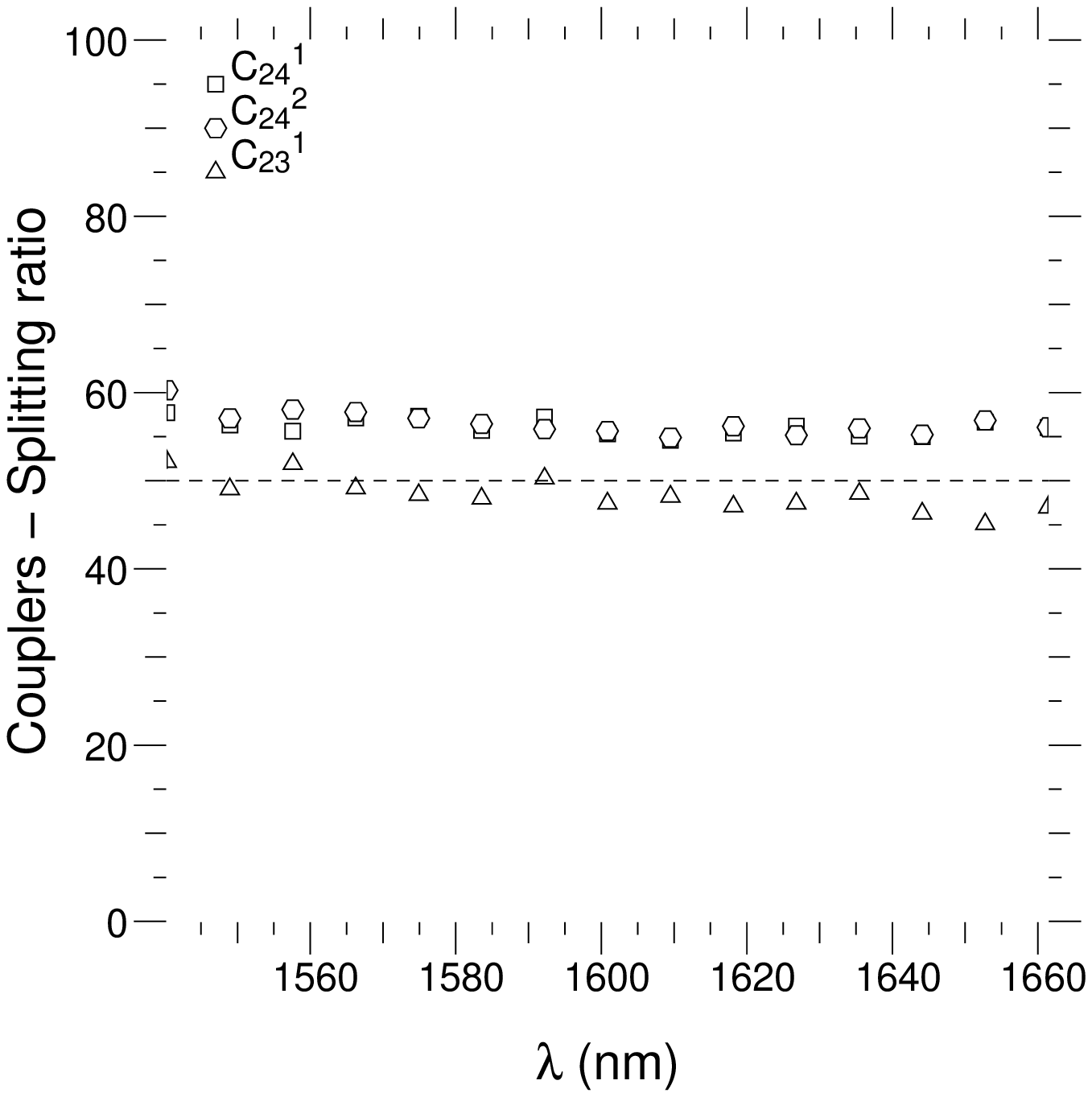}
     \end{tabular}
     \caption{\label{fig:cpl_lam}  Variation  of  coupler  splitting
       ratio with wavelength. Only one of the flux ratio coefficients is plotted,
       and  extreme  behaviors are  given~:  left,  when  the light  is
       injected  in the  first input;  right, for  an injection  in the
       second input.   The theoretical value  of 50\% is given  by the
       horizontal dashed line.}  
\end{center}
 \end{figure*}

 \begin{table*}
 \begin{center}
 \caption{\label{tab:coupl} Coupler splitting ratio.  The dispersion (rms) over 1024 measurements is
 0.2\%. The two first lines correspond to the wide
 band experiments  while the  third gives the  average value  and the
 variation amplitude over the wavelength range.}  
 \begin{tabular}{ c||c c  c c  c c  c c  c c  c c }
\hline
\hline
 C&  C$_{12}^{1}$  &  C$_{12}^{2}$  &  C$_{13}^{1}$  &  C$_{13}^{2}$  &
 C$_{14}^{1}$ & C$_{14}^{2}$& C$_{24}^{1}$ & C$_{24}^{2}$ & C$_{23}^{1}$ & C$_{23}^{2}$ & C$_{34}^{1}$ & C$_{34}^{2}$\\
 \hline
 Chip 1 & 54.1 & 59.6& 50.5 & 58.4 & 59.7 & 61.5 & 57.5 & 58.5 & 48.3 & 50.4 & 53.9 & 55.2\\

 Chip 2 & 52.0 & 58.9 & 55.2 & 55.7 & 58.3 & 56.3 & 56.3 & 51.2 & 52.6 &55.0 & 55.0 & 55.4 \\ 

avg / $\Delta  \lambda$ & 53.9/5.4 & 53.9/4.1 &  55.6/4.5 & 57.2/5.7 &
56.0/6.1  & 60.0/5.8 &  56.0/3.1 &  56.5/5.4 &  54.8/7.0 &  47.6/5.0 &
47.2/5.0 & 54.1/7.2 \\ 
 \end{tabular}
 \end{center}
 \end{table*}

\bigskip
\textit{Couplers:} 
Table~\ref{tab:coupl} gives  the flux splitting ratio given  by the 12
couplers, in  broad band, showing asymmetric  splitting, up to
61.5\%/38.5\%, similarly for the two beam combiners.  \\
With spectral dispersion, the obtained values are similar.
The variation of the 12 coupler splitting ratio with wavelength
is given in Fig.~\ref{fig:cpl_lam} (with only extreme behaviors). All
couplers show low chromaticity, with an average maximum variation of
about 5\%, with minimum and maximum values of about 3 and 7\%.

\vskip 0.1cm
Finally, although, in theory, all functions are identical, we notice some
slight variations from one  another. In broad band, for Chip1,
  the differences are up to 1.7\%, 2.7\% and 4\%, for the tricouplers,
  Y-junctions and couplers, respectively, while for Chip2, they are of
  0.9\%, 2.3\%, and 2.6\% for the same functions.

\begin{figure*}
  \begin{center}
    \begin{tabular}{cc}
      \includegraphics[width=0.35\textwidth]{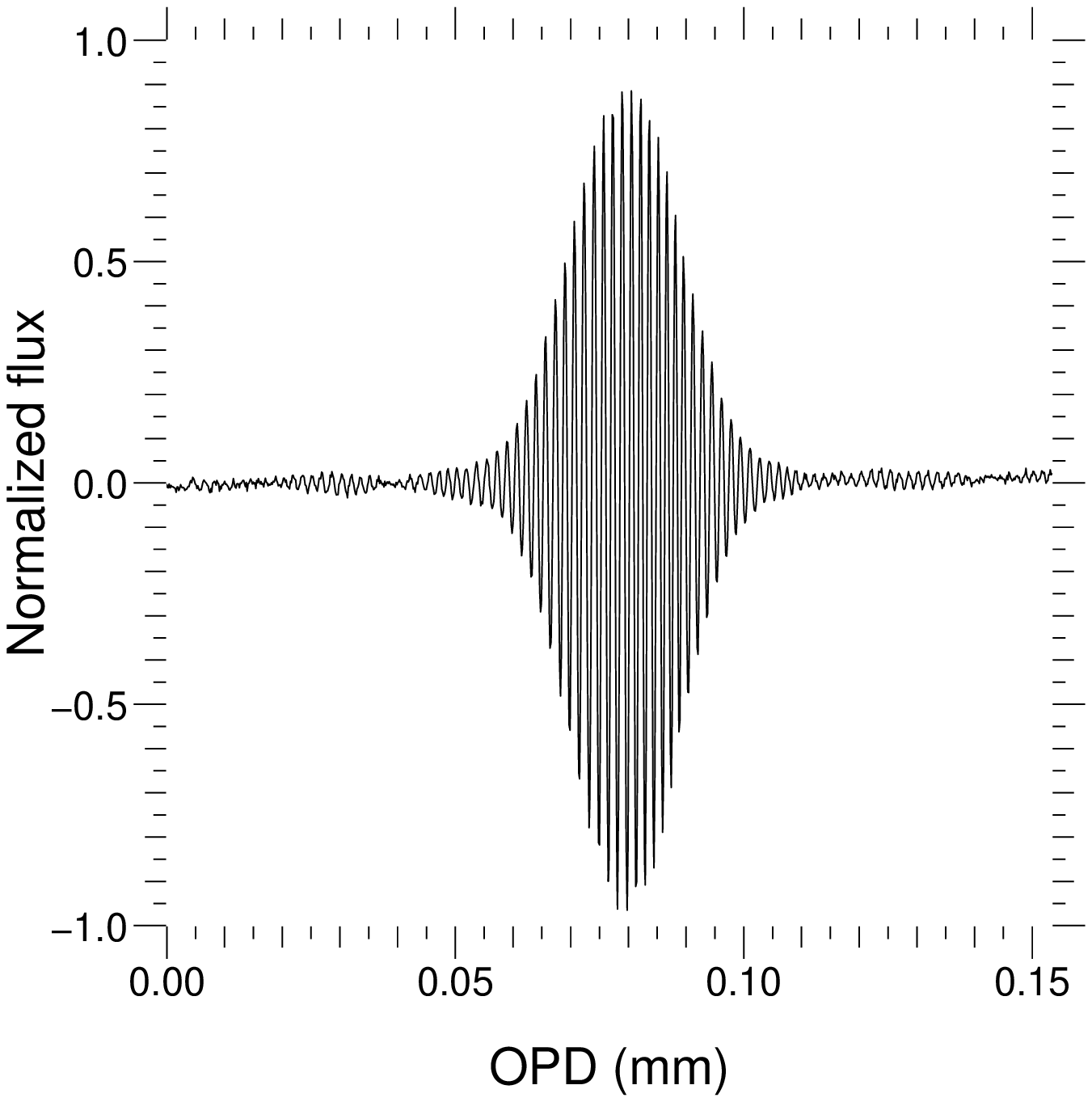} & 
      \includegraphics[width=0.32\textwidth]{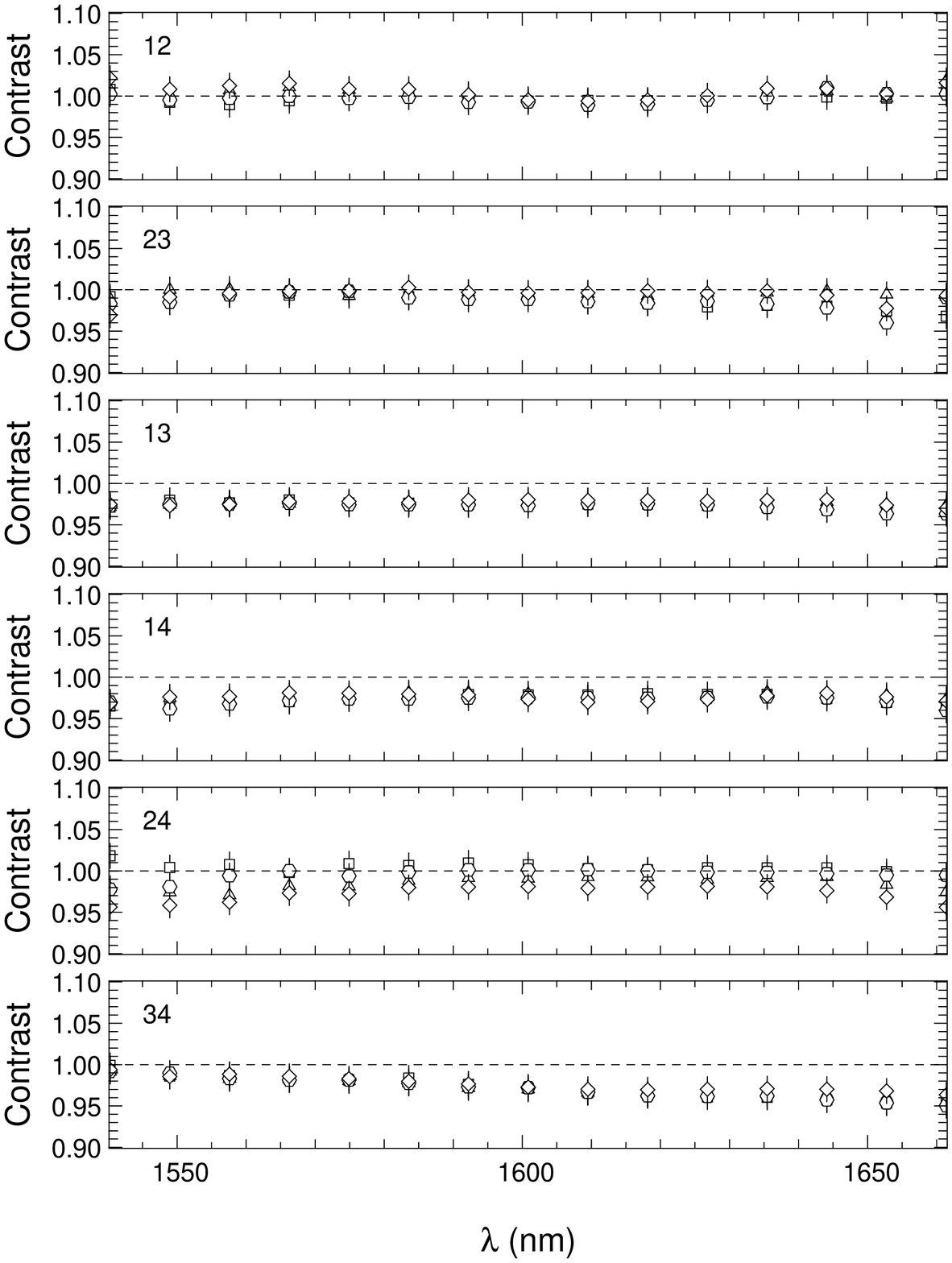}
\end{tabular}
    \caption{\label{fig:interf1} Left: an interferogram obtained with
      temporal OPD modulation with wide-band measurements.  Right: the
      variation  of instrumental  contrasts with  wavelength,  for all
      four outputs of each beam pair. 100\% contrasts are given by the
      horizontal dashed line.}  
\end{center}
\end{figure*}

 \begin{table*}
   \begin{center}
   \caption{\label{tab:ct} Cross-talk flux (in \%) determined when the
      light is injected in the 4 outputs successively. The two first lines
      report  results obtained  in  broad band  while  the third  line
      corresponds to the spectral dispersion case.  Error bars
      are estimated from the dispersion (rms).}  
 \begin{tabular}{c||c c c c} 
       \hline
       \hline
       Day & Injection in 1 & Injection in 2 & Injection in 3 & Injection in 4 \\
       \hline
       Chip 1 & 0.7$\pm$0.2 & 0.6$\pm$0.2 & 1.2$\pm$0.4 & 1.1$\pm$0.3\\

       Chip 2 & 0.4$\pm$0.2 & 0.3$\pm$0.2 & 0.3$\pm$0.2 & 0.2$\pm$0.2 \\

       avg / $\Delta \lambda$ & 0.3/0.03 & 0.4/0.03 & 0.9/1.5 & 0.5/0.6\\
     \end{tabular}
   \end{center}
 \end{table*}

\bigskip
\textit{Cross-talk:} 
On  the pixels  where one  should not  detect any  flux,  the measured
intensity gives the amount of undesired flux, \textit{i.e.} cross-talk
flux. It can be due to direct propagation into the substrate and
to light leak at the X-junctions level, where waveguides are crossing.  We estimate the
cross-talk  flux to  be  less  than 1.2\%$\pm$0.4\%  of  the  total
flux. Table~\ref{tab:ct}  gives the measured cross talk  flux when the
light is  injected in one  input at the  time.  The impact of  such an
amount  of cross-talk on  the photometric  quantities  is within
their error bars.  Also  included in Table~\ref{tab:ct} is the average
cross-talk over wavelength with its chromaticity.

\begin{table*}
\centering
\caption{\label{tab:c} Instrumental contrasts (in \%) obtained with wide band
  experiments (first  two lines)  and with spectral  dispersion (third
  line) with a statistical error of 1\%.}  
\begin{tabular}{c}
\begin{tabular}{ c||c c c c  c c  c c c c  c c }
\hline
\hline
Output &  $12^{1}$ & $12^{2}$ & $12^{3}$  & $12^{4}$ &
$23^{1}$   &  $23^{2}$   &  $13^{1}$   &   $13^{2}$  &
$13^{3}$ & $13^{4}$ &  $14^{1}$ &  $14^{2}$ \\
\hline
Chip 1 & 95$\pm$1 & 97$\pm$1 & 95$\pm$1 & 98$\pm$1 & 96$\pm$1 & 97$\pm$1 &94$\pm$1 & 96$\pm$1 & 94$\pm$1 & 95$\pm$1 & 96$\pm$1 & 98$\pm$1\\ 

Chip 2 & 90$\pm$2 & 90$\pm$2 & 92$\pm$2 & 88$\pm$3 & 85$\pm$3 & 93$\pm$2 & 83$\pm$5 & 88$\pm$4 & 85$\pm$4 & 88$\pm$3 & 88$\pm$4 & 94$\pm$1 \\

avg / $\Delta \lambda$ & 99/2.7 & 100/2.1 & 99/2.6 & 100/1.8 & 98/3.3 & 99/3.8 & 97/1.0 & 97/1.3 & 97/0.8 & 97/1.4 & 97/1.3 & 97/1.6 \\
\end{tabular}
\\
\\
\begin{tabular}{ c||c c  c c c c  c c  c c c c}
\hline
\hline
Output &  $14^{3}$ &  $14^{4}$ &  $24^{1}$  &  $24^{2}$ &
$24^{3}$ &  $24^{4}$ & $23^{3}$ & $23^{4}$ &  $34^{1}$ &  $34^{2}$ &  $34^{3}$ &  $34^{4}$ \\
\hline
Chip 1 & 97$\pm$1 & 97$\pm$1 & 96$\pm$1 & 97$\pm$1 & 97$\pm$1 & 97$\pm$1 & 97$\pm$1 & 97$\pm$1 & 96$\pm$1 & 97$\pm$1 & 97$\pm$1& 96$\pm$1\\

Chip 2 & 88$\pm$4 & 93$\pm$1 & 84$\pm$4 & 93$\pm$2 & 88$\pm$3 & 91$\pm$2 & 86$\pm$4 & 92$\pm$2 & 82$\pm$6 & 88$\pm$3 & 83$\pm$8 & 84$\pm$3\\

avg / $\Delta \lambda$ & 97/1.2 & 97/1.5 & 100/2.5 & 98/2.4 &
99/2.2 & 97/2.4 & 98/1.1 & 99/2.6 & 97/3.0 & 97/4.1 & 97/3.5 &   97/4.6 \\
\end{tabular}
\end{tabular}

\end{table*}

\subsection{Interferometric measurements}
\textit{Instrumental contrasts~:} 
Figure~\ref{fig:interf1}, left,  shows an example  of the interferograms
obtained with Chip~1 in broad band.  From these interferograms, we derive the
instrumental contrasts after calibrating for the photometric inbalance
between  interfering  beams.   Table~\ref{tab:c}  gives  the  measured
values, showing high contrast  values.  The minimum and maximum values
are  respectively  of  95\%$\pm$1\%  and 98\%$\pm$1\%  for  the  first
beam combiner,  and of 82\%$\pm$6\% and  94\%$\pm$1\% for the
second one. \\
These  results  show a  maximum  variation of  about  5\%  over a  day
timescale and 10\% from one day to another.  The measured contrasts on
the four phase shifted outputs are only slightly different (on average
2\%, up to  9\%).  The non-perfect linearity of  our detector can lead
to a bias of up to 5\% in visibility.  In the case of our experiments,
this  effect  could  not  be  reproduced  and  calibrated.  Therefore,
although  the statistical  errors can  be very  small  ($\sim$1\%), an
additional bias of 5\% affects the contrast values.\\
With spectral dispersion, the measured contrasts are very high
  (up  to 100\%) showing  very small  variations with  wavelength (see
  Figure~\ref{fig:interf1},   right).    The   maximum   and   minimum
  chromaticities,  among  all  24  outputs,  are of  4.6\%  and  0.8\%
  respectively, with an average of 2.3\%.

\begin{figure*}
  \begin{center}
    \begin{tabular}{cc}
      \includegraphics[width=0.33\textwidth]{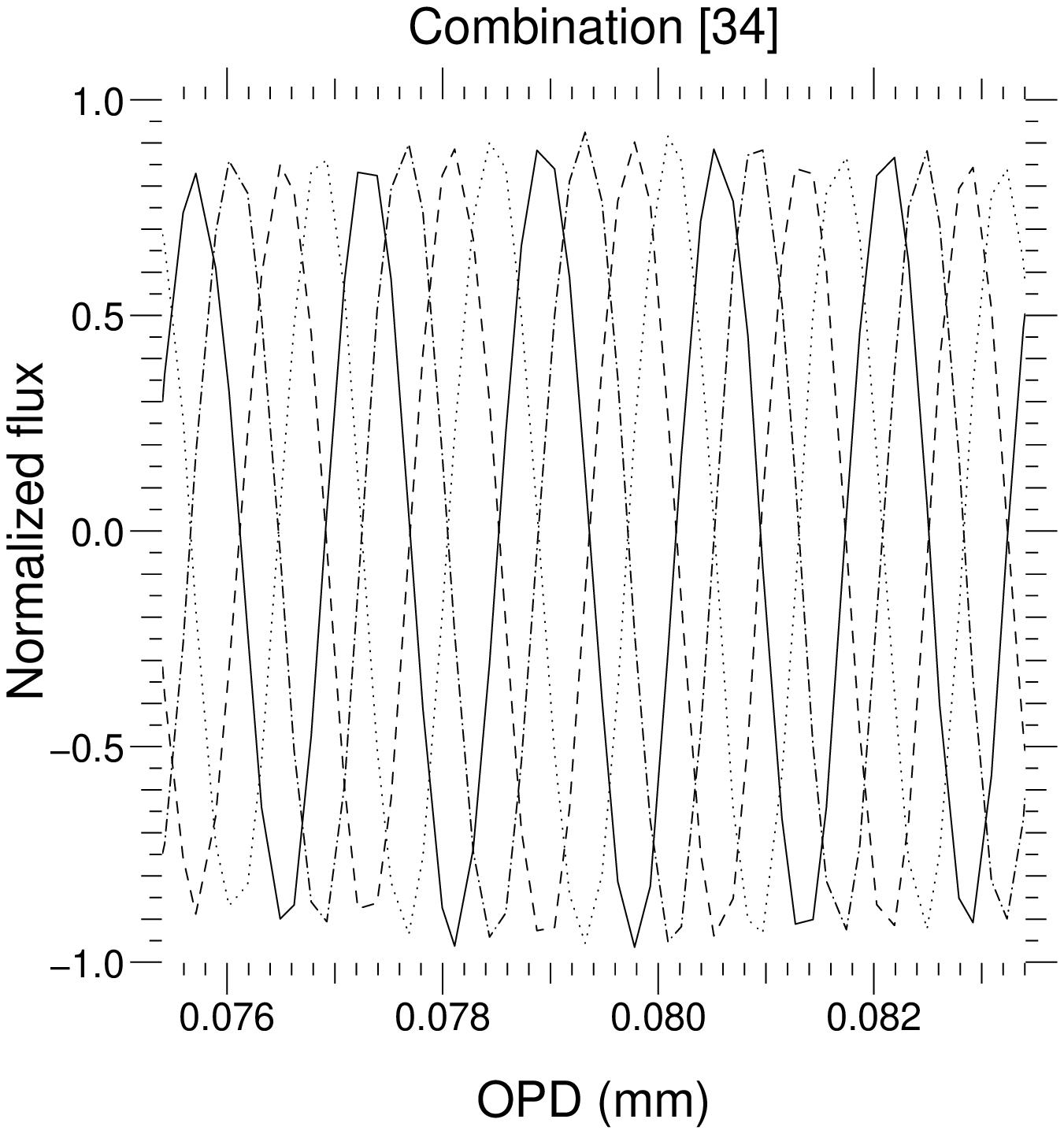} & 
      \includegraphics[width=0.33\textwidth]{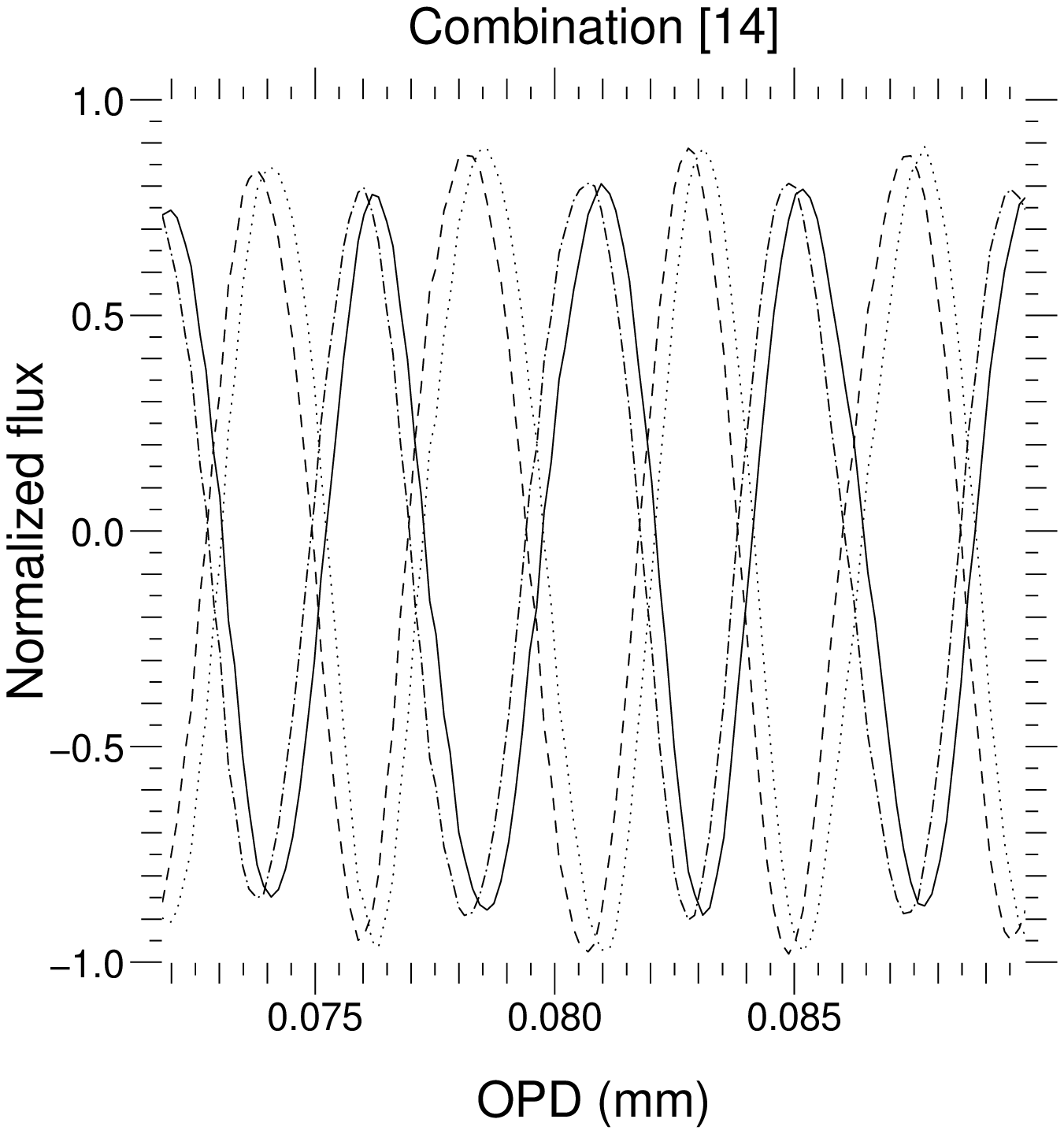} 
    \end{tabular}
    \caption{\label{fig:interf2}   The   phase-shifted  interferograms
      recorded  for the  4 outputs  (full, dashed,  dotted, dot-dashed
      lines) of the [34] beam  pair are close to quadrature (left). On
      the  contrary, the  4 outputs  of the  central  combination [14]
      produce     interferograms    only     slightly    phase-shifted
      ($\sim$26$^\circ$) (right).}  
\end{center}
\end{figure*}

\begin{figure*}
  \begin{center}
    \begin{tabular}{cc}
     \includegraphics[width=0.35\textwidth]{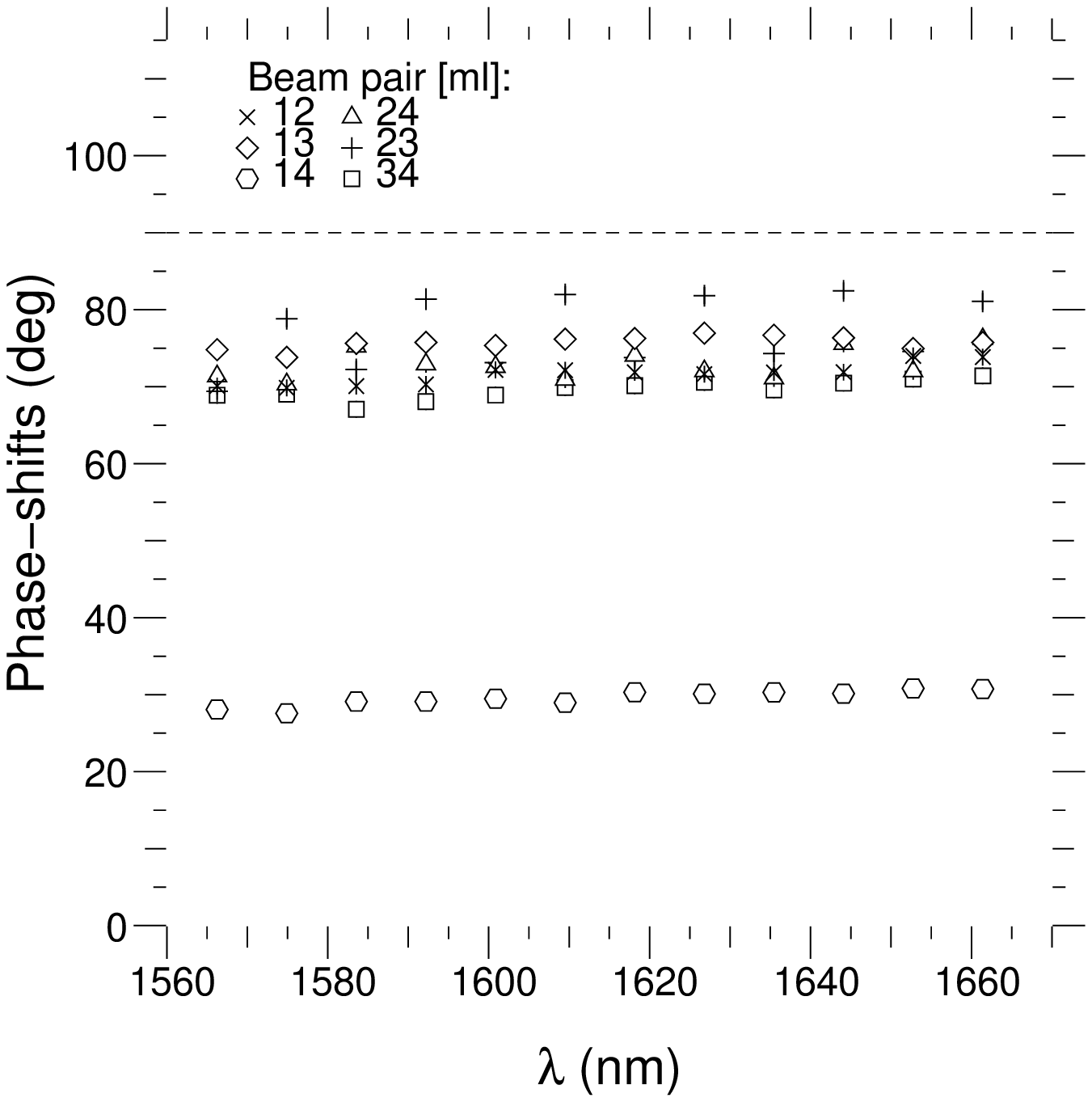}
      & 
      \includegraphics[width=0.35\textwidth]{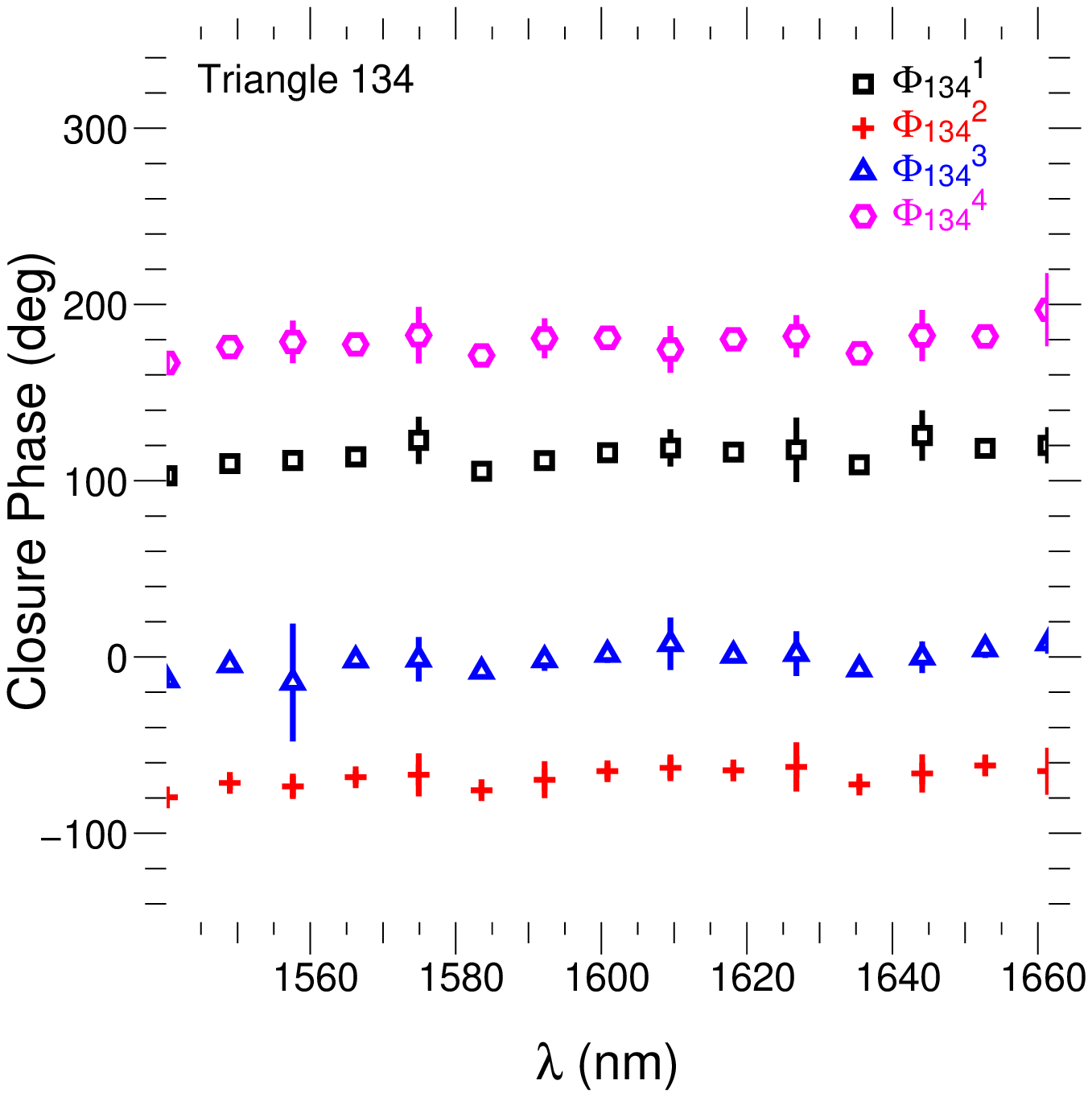}
 \end{tabular}
    \caption{\label{fig:dph_lam_cp} Left: wavelength dependence of the
phase-shifts   for  the  outputs   designed  to   be  in
      quadrature,     for      all     6     pairwise     combinations
      ([ml]).  Error  bars on  each  spectral  point are  of
      1$^\circ$. The theoretical value of 90$^\circ$ is given by the
      horizontal dashed  line.  Right: variation of  the closure phase
      over  the H  band, for  the triangle  [134]. The  four symbols
      correspond to the four outputs in quadrature.}  
\end{center}
\end{figure*} 

\begin{table*}
\begin{center}
\caption{\label{tab:deph}  Phase-shifts obtained  with  the wide  band
  experiments  (two  first  lines)  with errors  of  1.0$^\circ$  (the
  dispersion (rms) over half a day); The third line gives
the average values over the wavelength range.}  
\begin{tabular}{ c||c c c c c c}
\hline
\hline
Phase-Shifts & $\Phi_{12}$ ($^\circ$) & $\Phi_{13}$ ($^\circ$) & $\Phi_{14}$ ($^\circ$) &$\Phi_{24}$ ($^\circ$) & $\Phi_{23}$ ($^\circ$) & $\Phi_{34}$  ($^\circ$)\\
\hline
Chip 1 & 79.0 & 81.5 & 26.1 &79.4 &81.6 & 77.9 \\

Chip 2 & 62.5 & 87.3 & 32.9& 67.5 & 77.5 & 87.7 \\

avg / $\Delta \lambda$ & 71.6/4.1 & 75.7/3.1 & 29.5/3.2 & 72.8/5.8 &
77.1/13.0 & 69.6/4.3 \\
\end{tabular}
\end{center}
\end{table*}

\begin{table*}
\begin{center}
\caption{\label{tab:CP} 
Closure-phase measurements  for one  independent triangle, for  Chip 2
with spectral dispersion,  as well as amplitude of  the variation over
the H band range. Statistical errors are of 2.5$^\circ$.  } 
\begin{tabular}{c||c c c c}
\hline
\hline
Triangle [134] & $\Phi_{134}^{1}$ ($^\circ$)& $\Phi_{134}^{2}$ ($^\circ$)& $\Phi_{134}^{3}$ ($^\circ$)& $\Phi_{134}^{4}$ ($^\circ$)\\
\hline
avg / $\Delta\lambda$ & 114.6/22.8 & -68.2/18.1 & -1.9/22.2 & 179.0/30.1\\
\end{tabular}
\end{center}
\end{table*}

\begin{table}[t]
\begin{center}
\caption{\label{tab:phshpol}  Examples  of  phase-shifts  (in  degrees)
  measured in  the two  polarization states (P1,  P2), for the  6 beam
  pairs. Statistical errors are of 1$^\circ$.}  
\begin{tabular}{ c||c c c c c c}
\hline
\hline
Beam Pair & [12] & [13]  & [14] &[24] & [23] & [34]\\
\hline
P1 & 62 & 87 & 34 & 67 & 77 & 88 \\

P2 & 55 & 80 & 37 & 62 & 70 & 77 \\
\end{tabular}
\end{center}
\end{table}

\bigskip
\textit{Phase shifts:} 
Figure~\ref{fig:interf2}  shows 4  phase-shifted interferograms  in each
panel,  that  correspond to  the  intensity  of  the 4  phase-shifted
outputs for the interferometric couple [34].  These 4 outputs are in
different  phase states,  as  it can  be  seen, and  in these  specific
examples, in the left panel, the phase-shift is close to the expected value of quadrature.

Table~\ref{tab:deph} gives the values of the phase-shifts obtained for
the  outputs designed  to be  in quadrature.   For  the first
chip, on 5 of the 6 interferometric combinations, the phase shifts are
close    to   the    quadrature   (from    78$^\circ\pm$1$^\circ$   to
82$^\circ\pm$1$^\circ$).   For   the  sixth  phase-shifting  function,
corresponding to the central [14] combination, the measurement gives
about      26$^\circ\pm$1$^\circ$.     The     right      panel     of
Fig.~\ref{fig:interf2} corresponds to such a combination.  
The second chip gives different results, with phase shifts spanning a 
larger  range   of  values  (\textit{e.g}  62$^\circ\pm$1$^\circ$;
88$^\circ\pm$1$^\circ$).  The  function  corresponding to  the  [14]
combination   still   shows   a    much   smaller   value   of   about
33$^\circ\pm$1.0$^\circ$.  

Over a timescale of half a day, measured phase shifts show variations
of 1$^\circ$ at most, and from one day to another, a maximum variation
of 3$^\circ$.  

With  spectral   dispersion,  the  obtained   values  for  the
phase shifts are similar.  Figure~\ref{fig:dph_lam_cp} (left) and
Table~\ref{tab:deph} show the wavelength dependence of the measured
phase-shifts for the outputs expected to be in quadrature.  The
maximum chromaticity  is of 13$^\circ$ (for  the [23] combination,
as for the Y-junctions) and  the minimum is of 3.1$^\circ$. Over all
outputs, the average chromaticity is 5.6$^\circ$.

\bigskip
\textit{Closure phases:} 
Table~\ref{tab:CP} gives a set of independent instrumental
closure~phases measured for the 4 phase~shifted outputs of one
telescope triangle, with spectral dispersion.  The measured 
closure~phases have non-zero values which means that the beam combiner
functions themselves contribute to  the phase budget.  These terms can
result  from additional  OPD originating  in small  length differences
between waveguides or from the delay in reading the detector pixels. 
The phase relation between the various beam combiner outputs can be found
again in the closure-phase values.  In fact, the closure-phases differ by about 180$^\circ$ for couplers outputs in phase
opposition and similarly, the closure phases measured at outputs theoretically
in quadrature are different to the corresponding phase sum between the
telescopes (\textit{i.e.} $\varphi_{ml} + \varphi_{lj} - \varphi_{mj}$). 

Figure~\ref{fig:dph_lam_cp}  (right) and  Table~\ref{tab:CP}  give the
variation of  the instrumental closure~phases with  wavelength, in the
case of one triangle of telescopes ([134]).  The minimum variation over the
wavelength range is of 18$^\circ$ while the maximum is of 30$^\circ$.

\begin{table}[!t]
\begin{center}
  \caption{\label{tab:sepPola}  Contrasts  and  phase-shifts  obtained
      when the splitting of polarizations  is done before or after the
      combination, and when no splitting is achieved. Errors are statistical.}
  \begin{tabular}{c}
    \begin{tabular}{c||c c c c }
      \hline
      \hline
      Pixel & $12^{1}$ & $12^{2}$ & $12^{3}$ & $12^{4}$\\
      \hline 
      $<V>$ before & 83.7$\pm$0.3 & 83.6$\pm$0.3 & 81.3$\pm$0.3 & 81.7$\pm$0.3 \\

      $<V>$ after & 79.6$\pm$0.2 & 79.4$\pm$0.2 & 78.2$\pm$0.2 & 77.2$\pm$0.2 \\

      $<V>$ $\varnothing$ & 71.3$\pm$0.5 & 71.2$\pm$0.5 & 66.7$\pm$0.5& 66.9$\pm$0.5 \\
    \end{tabular}
    \\
    \\
    \begin{tabular}{ c||c c c }
      \hline
      \hline
      Polar splitting & before & after & $\varnothing$\\
      \hline
      Phase-shifts &  78.98$\pm$0.03 & 79.80$\pm$0.08 & 76.12$\pm$ 0.07\\
    \end{tabular}
  \end{tabular}

\end{center}
\end{table}

\bigskip
\textit{Polarization:} 
The two linear polarization directions are perpendicular to each other
  following the symmetry axis of the chip itself (\textit{i.e.} 
  within  the beam  combiner plane  (horizontal) and  parallel  to the
  light wavefront  (vertical)).  Such orientations are  defined at the
  time of  manufacturing the chip,  and were confirmed  by laboratory
  measurements.

In our study, disparities between measurements obtained on both 
linear  polarization  states  were  noticed.   Contrasts  and
phase-shifts can be up to, respectively, 
13\%  and 10$^\circ$  different,  meaning that  the two  polarizations
propagate differently.  Also, for one polarization state, instrumental
contrasts  are higher  than for  the second  one for  all beam
  combiner outputs \textit{except} for the ones related 
to the  central phase shifting function  ([14] combination).  Similarly,
phase shifts are closer to the quadrature for one polarization than the other for all outputs except the central ones
(Table~\ref{tab:phshpol}).   Therefore, with respect  to polarization,
one phase shifting function has a different behavior than the 5 others.

In  order  to  identify  possible problems  with  instrumental
  polarization and to determine the best instrumental set-up to 
  reduce the  visibility loss due  to birefringent fibers, a detailed
  study on a day timescale was done with three optical set-ups: the
first one  had polarizers \textit{before} injection  into the fibers
  and  beam combination (\textit{i.e.}  before the  telescope mounts);
  the second one had a Wollaston prism before imaging on the detector,
  \textit{after}  beam   combination;  and   the  third  one   had  no
  polarization splitting.  

We found that  the instrumental contrasts can drop by  10 to 15\% when
no   polarization   splitting   is  done   (Table~\ref{tab:sepPola}).
Slightly better contrasts ($\sim$3\%) were found when using 
polarizers before beam combination rather than after with a Wollaston
prism.    When  comparing   stability  for   all   configurations,  no
significant difference was found.

\section{Discussion}
\label{sec:dis}
The characterizations presented here shows that global properties
of the designed beam combiners are very satisfactory for a first prototype.
In this section, we discuss  how departures from the ideal case might affect
the performance.  

\vskip 0.2cm
\textit{Photometry:} The transmission of the beam combiners directly impacts the instrument sensitivity
and therefore,  constrains the limiting magnitude.   For these longest
and  most   complex  IO   beam  combiners  tested   today,  a
65\% transmission is satisfactory. 
An improved technology has allowed to reduce the losses with respect
to previous beam combiners.  For comparison, the IONIC3T/IOTA
H band beam combiners that were made with the first technology
and had only 3 Y-junctions and 3 couplers, presented a transmission of
60\%.  The  gain comes  essentially  from  improvement in  propagation
losses associated with a reduced beam combiner length.

We show evidence for small crosstalk photon leaks that lead to unwanted flux in the combining
cells. These potentially affect the photometry estimation and might introduce a small coherent
contribution not revealed in this study. However the measured values for both chips are mostly
inferior to 1\% of the total flux and have contributions smaller than the typical error bars of the
measurements.  The origin of this effect has been identified as coming from imperfect fiber/IO
coupling  (in the  experiment the  fibers are  not glued  unlike  in an
actual instrument)  and from  flux coupling in  the substrate  that is
partially guided. However, the 
latter  contribution  has  been  dramatically reduced  thanks  to  new
etching  technology  that  allows  each  waveguide  to  be  completely
isolated from the others and to not waste flux \citep{labeye06}.  As a
matter of fact,  our experiments using the old  technology showed flux
crosstalk of up to 8\%.  

The flux routing of the interfering beams, for each pair, needs to be as equal as possible,
although this is not a strong requirement. In fact, the instrumental visibility (and therefore the
SNR)  decreases   as  the  beam  fluxes  are   unbalanced.  For  these
beam  combiners, the  tricouplers equally  split the
flux in three (within the error bars), and the flux splitting ratios of
the Y-junctions are satisfactory. On the contrary, the couplers can be
as unbalanced as 60\%/40\% leading to a maximum contrast loss of 2\%.
These splitting  ratios were  estimated with the  assumption of
  ideal functions, which actually present a small loss of a few $\%$
\citep{labeye08}. These unbalanced ratios are  due to an error in the
design,  and new  prototypes are  being made  with  improved couplers.
Globally, the agreement of the simulated data with the experiments is very
satisfactory.

Finally, the  stability and repeatability  of all observables  are key
elements to  reach high  dynamics.  Photometric quantities  are stable
over a day timescale at the level of a percent, and are reproducible
from one  day to another with  a maximum variation of  3\%. We suspect
that the absence of a glued interface between the fibers and IO beam
combiners contributes to 
the  small errors (except  for the  closure phases  for which
  this effect is canceled  out). However, for a first demonstration,
the performances are good and the general flux 
routing is validated.  The  competitive transmission together with the
capacity  to use  all  photons  for coherent  combination  lead to  an
overall high sensitivity for such 4-beam phase-shifting beam combiners.

\vskip 0.2cm
\textit{Interferometry:} The instrumental contrast directly impacts the sensitivity of the
instrument and should therefore be as high as possible.  Here the beam combiners produce instrumental
contrasts with values always above 80\% in wide band, and reaching 100\% with spectral dispersion.
In broad band operation, the effects of differential dispersion due to unequal length of the fibers can lead
to a contrast loss of up to 20\% which explains the very high contrasts obtained with spectral
dispersion.   Statistically, the stability of the contrasts over a day
timescale is good (from 1\% to 4\%) but the reproducibility of such measurements onto another day
could not be validated due  to the non-linearity of our detector. This
affects the value with a bias of 5\%, making our results only upper limits.

The  phase shifting functions  were initially  designed to  sample the
coherence  in four  phase states  in quadrature  (the so  called ABCD
sampling). For the two tested beam combiners, the phase-shifting 
functions lead to 5 phase-shifts out of 6 in agreement with our expectations, while one of them (the
central one corresponding to the [14] combination) is far from 90$^\circ$.  We suspect an
inhomogeneity in the constitution of the silicon substrate at the position of this phase shifting
function. Of all 6 functions, this one is the most distant from the center of the beam combiner and is
located  next to  the edge  of  the chip.   This effect  could
  explain  the measurements  obtained  on both  chips  since they  were
  located next to each other on the same wafer. This could also result from a relaxation of
the stresses  on the  silica during  the chip cut.  An error in the
design  has  been  ruled out.   This  is  the  first time  that  these
functions have been tested and while still 
not completely controled, these  results are encouraging and show that
the use of phase shifting functions in IO beam combiners is promising.
In addition,  phase shifts are stable  on a timescale of  a day, within
1$^\circ$ and vary up to 3$^\circ$ from one day to another.  

The incidence of this departure from phase quadrature on the final complex visibility SNR estimation
cannot be quantified without a proper calculation. It should be seen as a reduction in the
instrumental response. In the limiting case where the phase shift is 0, one cannot retrieve
unambiguously  the complex visibility  information. However  since all
but one phase shift have values close to quadrature, we believe this
validates the concept.

Finally, the measured closure phases are not equal to zero. Specific phase contributions of each
function, that result from non-perfectly symmetric pathways, are unknown.  When observing a
scientific  target,   calibrating  with   a  point  source   (that  is
centro-symmetric and supposedly leads to a zero closure-phase) should allow us to remove the instrumental
contribution.  It  should  be   expected  that  given  the  remarkable
stability  of such  beam combiners  \citep{berger00}  the instrumental
response should be very well calibrated.  

\vskip 0.2cm
\textit{Chromaticity:} The importance of the chromaticity of the functions directly depends on the
spectral resolution used in the instrument.

On all photometric coefficients, the average chromaticity can go from about 12\% to 80\%.
Individual functions show varieties of chromaticity, with variations for tricouplers and Y-junctions
from 3\% to 28\% while couplers can be considered as achromatic. The Y-junctions and couplers for
the [23] combination are more chromatic than the others. More details will be
given in  a following  paper (Labeye et  al. 2009, in  prep.). However
photometric  calibration in  the presence  of dispersion  should solve
this issue. With correction for photometric effects, the
obtained contrasts show very little chromaticity, with an average maximum variation of 4.6\% over
the range of H band. Phase~shifts show chromatic variations of 5.6$^\circ$ on average, the maximum
variation corresponding to the [23] combination as for individual functions.  The closure phases
show strong variation with wavelength, up to 30$^\circ$, which results in the sum of all chromatic
effects for the three telescopes in the closure relation.

We can anticipate that in cases where spectral resolution is low the effective wavelength might be
affected and a proper wavelength calibration should be considered together with a proper stellar
calibrator choice. In addition, in order to limit potential biases, particular care should be
taken to specify the alignment accuracy between the beam combiner, spectrograph and detector pixels.

\vskip 0.2cm
\textit{Polarization:} Birefringence control is a critical part of a guided optics instrument.
In our  experiment, we  used highly birefringent  fibers which  have a
well defined polarization axis but,  in turn, require specific care on
how the polarization state is modified along the propagation. 

Our study first shows that all estimated quantities (photometry, instrumental contrast, phase
shift, closure phase) can be different depending on the polarization state. We have shown in
particular that the behavior of the central [14] combination is different from the other
phase-shifting functions. This confirms the suspected marginal behavior of this combination, maybe
originating from an inhomogeneity in the substrate for both chips. \\
Besides, our work showed that it is necessary to split the polarization states before or after the
beam combinations, or to actively control the phase shift between the two linear states to avoid
contrast loss.  The use of birefringent fibers and waveguides forces the phase velocity of
the two polarization states to be different and results in two shifted interferograms. When these
two are superimposed, it leads to a single low contrast interferogram.  In the case of a sensitive
imaging instrument, splitting after the beam combination is recommended since the use of polarizers
before the combination would lose half of the useful photons.  As far as stability is concerned,
without splitting, the differential phase shifting of the polarization
states inside the beam combiner
would lead to varying results.   The new prototypes are being designed
with special care given to
this problem.

\vskip 0.2cm
\textit{On-sky operating modes:} By allowing all the complex coherent factors to be measured in one
single detector frame readout, these beam combiners offer two observational modes, depending on the
stability of the fringes (\textit{i.e.} on the atmospheric conditions or on the availability of a
fringe tracker and its performance).  If the fringes are stabilized to
better than a fraction of  the wavelength, a long coherent integration
of the flux on each pixel is possible (\textit{i.e} 
coherencing  mode), highly  increasing  the SNR  compared to  temporal
encoding.  Otherwise, by varying the
OPD,  one  will access  4  phase-shifted  interferograms  on which  to
estimate  the interferometric  observables.  The  latter mode  is also
well suited for laboratory measurements and calibration.  

\vskip 0.2cm
\textit{Precision interferometry and data reduction:} 

We have shown that the described beam combiners present performance well suited for astronomical beam
combination in a four  telescope imaging interferometer. Our extensive
laboratory characterization shows that on-sky performance, in terms of
precision,  should  be  comparable  to  what has  been  achieved  with
IONIC-VINCI/VLTI, IONIC3/IOTA or FLUOR/IOTA-CHARA.  However, an interesting number of astrophysical problems will soon be more demanding (e.g. debris disk,
hot  Jupiter  detection). In  that  case  it  is important  to  better
characterize calibration issues
and tackle all imperfections that could be introduced by the beam combiner. Such work is justified by
the tremendous stability of the beam combiner that has been revealed by numerous industrial
developments. Therefore, a proper calibration of the beam combiner  should allow systematics biases
to be removed. What we propose is to use the so-called 'visibility to pixel matrix' calibration
in order to carry out a global inversion of the matrix that links the measured intensities with the
properties  of the  scientific object.  Such  a method,  that will  be
detailed in a coming paper allows all instrumental contributions to be
extracted.  We briefly discuss here the philosophy of this data
reduction.  By developing the cosine, Eq.~(\ref{eq:interf}) can be written in such a
way that instrumental terms are separated from the object contribution
(V$_{ml}^{obj}$,   $\varphi_{ml}^{obj}$,   or   similarly   the   complex
visibility $\mathbb{V}$), leading to a system of linear equations that
includes the instrumental matrix (V2PM, ~see Sect.~\ref{sec:tech}):
$$ \mathbb{I}
= 
\mathrm{V2PM} *
\mathbb{V} $$

The complex  visibility of the  \textit{scientific source} will be  obtained by
inverting  the system.  A  full characterization  of  the V2PM  matrix
should be first done with internal calibration procedures. This
method allows all instrumental effects to be included in a single matrix
taking into account all crossing terms due to \textit{unideal behavior}. 


\section{Conclusion}
We have presented a laboratory characterization, in the H band, of an integrated optics beam combiner
dedicated to the combination of a four telescope interferometer.  It uses a novel 'pairwise static
ABCD' combination scheme which  optimizes the extraction of coherence
information: visibility amplitudes, phases and closure phases. Our measurements show that, although complex, the flux routing
inside the beam combiner is efficient and that the global throughput is
competitive (65\%). In
particular the comparison with simulated performance of several building block functions is very
satisfactory.   The instrumental  contrasts are  high  ($\ge$80\%) and
stable showing maximum variations of a few $\%$ within a day.  The
instrumental closure phases are different from zero but stable and probably caused by small internal optical path differences.  The specificity of
these beam combiners,  which is to produce four  phase shifted outputs
to sample the coherence information (for each baseline) in one integration, has been validated. Finally, the global chromaticity of the
beam combiner has required specific optimizations that are non-standard   with   respect    to   standard   telecommunication   IO
functions. IO chips  used in  telecommunications  are usually
  required to show a flat response on short bandpasses ($\Delta\lambda
  \approx  $20-30nm) while  here  we have  extended  the couplers  and
  tricouplers  and  phase  shifter  response  to  more  than  $\approx
  150~nm$ around $1.6 \mu\rm{m}$ \citep{labeye08}.  These results on a first prototype validate the feasibility of a 
'pairwise static ABCD' combination scheme and its suitability for an
  interferometric imaging instrument.  

Most of technological building blocks are now defined.  We are working
on the definitive version that will 
be included inside Gravity and VSI. This will require a number of
technological improvements and innovations. In particular, the
phase-shifting function will be ameliorated to get closer to the
nominal $90^{\circ}$ phase shift.  The throughput will be improved
with an optimized routing that will reduce the global propagation
losses.  We  are currently extending  the demonstration, using
  the same silica on silicon technology, to the K band, as required by
  Gravity and  VSI. It  is expected that,  while less  transmissive in
  this band,  the short propagation distances inside  the combiner will
  lead to  acceptable global losses.  Finally, we will  explore how
this combination concept can be extended to a six-way beam combiner and fit VSI requirements.

\begin{acknowledgements}
  This  work   was  financially   supported  by  CNES,   CNRS,  ASHRA,
  Universit\'e  Joseph Fourier  and Agence  Nationale de  la Recherche
  grant ANR-06-BLAN-0421.  We thank J-B.~Le~Bouquin, and 
  S.~Lacour  for  fruitful discussions.  We  acknowledge the  referee,
  Markus Schoeller, for his careful  reading of the manuscript and for
  thoughtful suggestions that improved its clarity.
\end{acknowledgements}

\nocite{berger08}

\bibliographystyle{aa}
\bibliography{paper_4tabcd}

\end{document}